\documentclass[12pt]{article}
\input epsf
\usepackage{graphicx}
\usepackage{bm}
\usepackage{latexsym}

\newcommand{\bea}{\begin{eqnarray}}
\newcommand{\eea}{\end{eqnarray}}
\newcommand{\be}{\begin{equation}}
\newcommand{\ee}{\end{equation}}
\newcommand{\pa}{\partial}
\newcommand{\nn}{\nonumber \\}
\newcommand{\e}{\epsilon}
\newcommand{\ve}{\varepsilon}
\newcommand{\w}{\omega}
\newcommand{\Tr}{\mbox{Tr}}
\newcommand{\tr}{\mbox{tr}}
\newcommand{\hH}{\hat{H}}
\newcommand{\hPhi}{\hat{\Phi}}
\newcommand{\hPi}{\hat{\Pi}}
\newcommand{\tPhi}{\tilde{\Phi}}
\newcommand{\tD}{\tilde{D}}

\makeatletter

\@addtoreset{equation}{section}
\makeatother

\def\href#1#2{#2}

\begin{document}

\begin{titlepage}
\vspace*{18mm}

\begin{center}
{\LARGE Lectures On AdS-CFT}\\
\vspace*{5mm}
{\LARGE At Weak 't Hooft Coupling}\\
\vspace*{5mm}
{\LARGE At Finite Temperature}\\
\vspace*{15mm}
{FURUUCHI \ Kazuyuki}\\
\vspace*{4mm}
{\sl Harish-Chandra Research Institute}\\
{\sl Chhatnag Road, Jhusi, Allahabad 211 019, India}\footnote{%
Address after Aug. 2007 : National Center for Theoretical Sciences,
National Tsing-Hua University, Hsinchu 30013, Taiwan,
R.O.C. ; E-mail: {furuuchi@phys.cts.nthu.edu.tw}}\\
{\tt furuuchi@mri.ernet.in}
\vspace*{4mm}
\begin{abstract}
This is an introductory lecture note 
aiming at providing an overview of
the AdS-CFT correspondence 
at weak 't Hooft coupling 
at finite temperature.
The first aim of this note
is to describe
the equivalence
of three interesting
thermodynamical phenomena in theoretical physics,
namely,
Hawking-Page transition to black hole geometry,
deconfinement transition in gauge theories,
and vortex condensation on string worldsheets.
The Hawking-Page transition 
and the deconfinement transition
in weakly coupled gauge theories
are briefly reviewed.
Emphasis is on the study of
't Hooft-Feynman diagrams in 
the large
$N$ gauge theories, which are 
supposed to describe  
closed string worldsheets
and probe the above equivalence.
Nature of the
't Hooft-Feynman diagrams
at finite temperature is analyzed,
both in the Euclidean signature
(the imaginary time formalism) and 
in the Lorentzian signature (the real time formalism).
The second aim of this note 
is to give an introduction to
the real time formalism
applied to AdS-CFT.
\end{abstract}

\end{center}

\setcounter{footnote}{0}
\end{titlepage}


\tableofcontents

\newpage
\section{Introduction}

The AdS-CFT correspondence
\cite{Maldacena:1997re}\footnote{%
See \cite{Aharony:1999ti} for a comprehensive 
introduction 
to the AdS-CFT correspondence.
We will use the terminology 
``AdS-CFT correspondence"
in a broader sense, 
for the duality 
between
closed string theory on
asymptotically AdS space
and field theory which approaches 
conformal fixed point at UV.
The most concrete example in mind is the
duality between
closed string theory on AdS$_5 \times S^5$
and ${\cal N}=4$ super Yang-Mills theory.}
has been providing us important insights
into 
strongly coupled gauge theories
and quantum gravity.
In particular,
the mysterious 
thermodynamical nature of black holes
has begun to be understood
by identifying it with the
dual boundary CFT at finite temperature.

If one deduce the
AdS-CFT correspondence
for four-dimensional CFT case
from the near horizon limit of
D3-branes,
one obtains the relation
between the radius $\mbox{R}_{AdS}$
of the AdS space
and the 't Hooft coupling
$\lambda \equiv g_{YM}^2 N$ 
in the boundary Yang-Mills theory as
\bea
\label{Rlambda}
\mbox{R}_{AdS}^4 \propto g_{YM}^2 N\, l_s^4 \, ,
\eea
where $l_s$ is the string length scale.
In the first stage of its theoretical 
developments,
AdS-CFT correspondence was mainly analyzed
in the strong 't Hooft coupling region,
where 
from the relation (\ref{Rlambda})
the curvature radius of the
AdS$_5$ is small and the supergravity approximation 
is valid.

Recently, weak 't Hooft coupling region
also began to attract wide interests.
In this case the curvature radius of 
the AdS$_5$ is around string scale,
and the supergravity approximation is not
valid.
And we do not yet know
how to quantize the string worldsheet theory 
on AdS space with R-R fluxes.\footnote{%
R-R fluxes should be there if we obtain
the AdS-CFT correspondence from
the near horizon limit of D3-branes.}
However, the Yang-Mills side is weakly coupled and
perturbative calculation is reliable.
Therefore, strategy in this region should be to 
extract the information of closed string theory
in highly curved background
from perturbative Yang-Mills analysis.%
\footnote{%
Since the relation (\ref{Rlambda})
was obtained from the supergravity solution,
we expect that it can be modified 
at weak 't Hooft coupling region
where the curvature of the geometry
reaches string scale.
Typically, we do not trust the relation
(\ref{Rlambda}) near $\lambda = 0$,
i.e. we do not assume that
the closed string dual of
a free Yang-Mills theory 
lives on the zero-radius (singular) AdS space.}

We are particularly motivated
by the two recent developments
in the weak 't Hooft coupling region.
One is the
deconfinement transition 
at weak 't Hooft coupling
in the large $N$ gauge theories on $S^3$
\cite{Sundborg:1999ue,Polyakov:2001af,%
Aharony:2003sx,Aharony:2005bq}.\footnote{%
We are benefitted from S. Minwalla's
brilliant lectures on their works.}
The other is the
Gopakumar's proposal for
how to reorganize 't Hooft-Feynman diagrams
into closed string amplitudes 
\cite{Gopakumar:2003ns,Gopakumar:2004qb,%
Gopakumar:2005fx,Gopakumar:2004ys,Aharony:2006th,%
David:2006qc,Yaakov:2006ce}.\footnote{See the references therein and in \cite{Furuuchi:2005qm}
for other approaches to obtain
closed string amplitudes
from perturbative Yang-Mills theories.}
This progress
motivates us to seriously
consider 't Hooft-Feynman diagrams
in the context of AdS-CFT.
Based on these two developments,
we have studied the 
properties of
't Hooft-Feynman diagrams
at finite temperature
in \cite{Furuuchi:2005qm,Furuuchi:2005zp}.\footnote{%
R. Gopakumar first suggested us to pursue his project
in the finite temperature case.
We would like to 
thank him for the collaboration 
in the early stage of \cite{Furuuchi:2005qm}
as well as many insightful suggestions.}
This lecture note is based on
these results, but
we also tried to incorporate
our new understandings.

In this note,
we'd like to give an 
overview
of the recent developments
in the AdS-CFT correspondence in the 
weak 't Hooft coupling region at finite temperature.
Section \ref{HP} and \ref{DecW}
are intended to provide sufficient
materials for reading the later sections.
References 
are provided for further learning.
In section \ref{HP},
after a brief introduction to the 
Euclidean path integral gravity,
we review
the Hawking-Page transition to black hole geometry,
which was later identified 
with the deconfinement transition 
in the dual Yang-Mills theory via 
the AdS-CFT correspondence.
Since the analysis is based on 
the Einstein-Hilbert action,
it is valid in the
strong 't Hooft coupling region
according to (\ref{Rlambda}).
However, the existence of the transition
seems robust and we expect
such gravitational transition 
with appropriate string corrections persists
up to the weak 't Hooft coupling region.
In section \ref{DecW},
we review the
deconfinement phase transition
of weakly coupled large $N$ gauge theories.
Section \ref{EtH} and section \ref{LtH}
treat the main subject of this note.
In section \ref{EtH},
we study
't Hooft-Feynman diagrams
at finite temperature in the imaginary time formalism.
How to incorporate the result of section \ref{DecW},
namely the effect of the phase being in 
confinement phase
or deconfinement phase, is described.
This leads to the picture
that 
the Hawking-Page transition to
black hole geometry,
the deconfinement transition,
and vortex condensation on the 
closed string worldsheets
are all equivalent.
Making
this picture clear 
is the first main aim of this note.
In section \ref{LtH},
we first review the real time formalism of
finite temperature theory
applied to the gauge field theory.
Then we describe
how to incorporate the effect of the
confinement/deconfinement 
phase background
into the 't Hooft-Feynman diagrams in this case.
The real time formalism corresponds to the 
Lorentzian signature geometry in the dual bulk theory,
where the real problems in black hole physics reside.
Giving an introduction to
the real time formalism and 
explaining its application in the 
AdS-CFT correspondence
is the second main objective of this note.


Our underlying attitude 
in the weak 't Hooft coupling region
will be that
rather than comparing the
results in two sides on an equal footing,
we put more weight on the Yang-Mills side.
We will mostly start from the perturbative 
Yang-Mills theory
and try to read off or
``define" the concepts in the bulk.

\section{Hawking-Page Transition and Deconfinement
at Strong Coupling}\label{HP}

In this section
we briefly review the
Hawking-Page transition \cite{Hawking:1982dh},
a thermal phase transition to a
black hole geometry in
asymptotically AdS space.
This was identified
as the dual description of the
deconfinement transition
in Yang-Mills theories
through the AdS-CFT correspondence
in \cite{Witten:1998qj,Witten:1998zw}.

\subsection{Euclidean path integral gravity}

The partitition function for 
canonical ensamble in a
quantum field theory
is given by
\bea
Z=\Tr \left\{ e^{-\beta \hat{H}(\hat{\phi},\hat{\pi})}
\right\} ,
\label{CE}
\eea
where
$\Tr$ is the trace in the Hilbert space 
and $\hat{H}$ is the Hamiltonian 
of the system of interest.
$\beta$ is the inverse temperature.
$\hat{\phi}$ is a field operator
and $\hat{\pi}$ is its conjugate momentum.
By following the usual steps
from canonical quantization to 
path integral,
we can rewrite (\ref{CE}) as
\bea
Z = \int {\cal D}\phi \, 
e^{- \int_0^\beta d\tau {\cal L}(\phi)} \, ,
\label{Epath}
\eea
where $\phi$ 
denotes
the 
path integral counterpart of
the field operator appearing in the (\ref{CE}),
and ${\cal L}$ is the Lagrangian.%
\footnote{We have
assumed that the Hamiltonian 
depends on the canonical momenta 
in a standard manner and they have been
integrated out.}
Since 
compared with the usual time translation operator
$e^{i \hat{H}t}$, 
$\beta$ in 
$e^{-\beta \hat{H}}$
can be seen as imaginary time direction,
this formalism is called 
imaginary time formalism.
It is also called 
Euclidean time formalism
because the rotation from
the real time to the imaginary time 
brings one from the Lorenzian time signature to
the Euclidean time signature.
In rewriting (\ref{CE}) to (\ref{Epath}),
it follows that
bosonic fields $\phi_B$ 
obey the periodic boundary condition
\bea
\phi_B(\tau + \beta) = \phi_B(\tau),
\eea
and fermionic fields 
$\phi_F$ obey the anti-periodic boundary condition
\bea
\phi_F(\tau + \beta) =  - \phi_F(\tau).
\eea

In the case of gravity,
we do not have satisfactory
formalism for 
canonical quantization yet.
However, we may just assume its existence and
start from the Euclidean path integral
(\ref{Epath}),
and take the classical approximation
\cite{Gibbons:1976ue}.
Then we obtain
\bea
Z \simeq e^{-I(\phi_{cl})},
\eea
where $\phi_{cl}$
is a classical solution 
of the equation of motion
with appropriate conditions 
imposed by the system of interests,
e.g. static, spherically symmetric and so on.\footnote{When
there are multiple solutions,
we sum over contributions from all those,
or may take the absolute minimum
as the leading contribution.}

\subsection{Hawking-Page transition}

Now we use the Euclidean path integral
formalism 
in the case of the 
asymptotically AdS space.\footnote{%
Here we are following the $n=4$ case of
section 2.4 of \cite{Witten:1998zw}.}
The model action we study here is
\bea
I
=
-\frac{1}{16\pi G_N}
\int d^5 x \sqrt{g} 
\left\{
R + \frac{12}{b^2}
\right\}.
\label{Igra}
\eea
Here, $G_N$ is the five-dimensional
Newton's constant.
In the case of the near horizon
geometry of D3-branes,
constant flux plays the role of
the cosmological constant in (\ref{Igra}),
and we get the similar results.
For a solution of the equation of motion,
$R_{\mu\nu}-\frac{1}{2}R g_{\mu\nu} 
-\frac{1}{2} \frac{12}{b^2} g_{\mu\nu} 
= 0$,
the value of the action becomes
\bea
I = \frac{1}{2\pi G_N b^2} \int d^5 x \sqrt{g},
\eea
that is, the volume of the space-time
times $\frac{1}{2\pi G_N b^2}$.%
\footnote{The action additionally has
a surface term \cite{Gibbons:1976ue}, 
but this vanishes
for the AdS-Schwarzschild black hole solution
because the difference between it and the AdS space
vanishes too rapidly at infinity \cite{Hawking:1982dh}.}
The AdS metric is a solution to
the equation.
In the global coordinates,
it is given by
\bea
ds^2 
=
\left(1+\frac{r^2}{b^2}\right) d\tau^2
+
\left(1+\frac{r^2}{b^2}\right)^{-1} dr^2
+ r^2 d\Omega_3^2 ,
\label{metAdS}
\eea
where $d\Omega_3^2$ is a usual
metric on $S^3$.
The $\tau$ direction is periodically
identified to discuss canonical ensamble
(the thermal circle).
To discuss canonical emsamble in gravity,
one first needs to choose which direction
to be the Euclidean time,
or equivalently should choose the Hamiltonian.
Our choice is to use the $\tau$ direction
in the coordinate (\ref{metAdS}).
With this choice, 
the asymptotic boundary at 
($r \rightarrow \infty$)
becomes $S^1\times S^3$ 
(up to a diverging overall factor which we rescale).
Therefore, the dual Yang-Mills theory lives
on the spatial manifold $S^3$.
We are interested in the
static spherically symmetric configurations.
Above certain temperature which will be described shortly,
there is another solution.
It is the
AdS-Schwarzschild black hole geometry
whose metric is given by
\bea
ds^2 
=
\left(1+\frac{r^2}{b^2} -\frac{\w_4 M}{r^2}\right) d\tau^2
+
\left(1+\frac{r^2}{b^2} -\frac{\w_4 M}{r^2}\right)^{-1} dr^2
+ r^2 d\Omega_3^2 ,
\label{metAdSBH}
\eea
where 
\bea
\w_4 \equiv  \frac{16\pi G_N}{3 \mbox{Vol}(S^3)},
\eea
and $M$ is the mass of the black hole.
In the Euclidean time signature,
the space-time is restricted to the region
$r \geq r_+$, where $r_+$ is the largest
solution to the equation
\bea
1+\frac{r^2}{b^2} -\frac{\w_4 M}{r^2} =0.
\eea
In the Lorentzian signature counterpart,
$r_+$ becomes the location of the 
black hole event horizon.
Therefore, the Euclidean geometry only covers
the region outside the horizon.
In the Euclidean path integral gravity,
temperature of the black hole is determined
from the requirement
that there should be no
conical singularity at the horizon $r=r_+$.
This determines
the period $\beta_0$ of the $\tau$ coordinate to be
\bea
\beta_0 = 
\frac{2\pi b^2 r_+}{2r_+^2+b^2}.
\label{b0}
\eea
From the AdS-CFT point of view,
a geometry with a conical singularity
is not exactly the saddle point
for a given temperature
and hence excluded.

The classical action
for the above solutions,
or the space-time volume which is proportional to it, 
is infinite so we need to regularize it.
We put a cut-off $R$
in the radial direction $r$.
Then, the volume
of the AdS space is given by
\bea
V_1(R) 
= 
\int_0^{\beta'} d\tau
\int_0^R dr \int_{S^3} r^3.
\eea
And for the AdS-Schwarzschild black hole, it is
\bea
V_2(R)
= 
\int_0^{\beta_0} d\tau
\int_{r_+}^R dr \int_{S^3} r^3.
\eea
We are interested only in the difference
of the two actions.
To compare the two, we 
match the physical circumference 
of the Euclidean time direction at $R$.
This determines $\beta'$ as
\bea
\beta' \sqrt{\frac{R^2}{b^2}+1}
=
\beta_0 \sqrt{\frac{R^2}{b^2}+1-\frac{\w_4 M}{R^2}}.
\eea
Then, the difference of the action
for the two solutions
is calculated as
\bea
I
=
\frac{1}{2\pi G_N b^2}
\lim_{R\rightarrow \infty}
(V_2-V_1)
=
\frac{\mbox{Vol}(S^3)(b^2 r_+^3 - r_+^5)}{4G_N(4 r_+^2 + 2 b^2)}.
\eea
The action becomes negative at
$r_+ \geq b$, or equivalently
$\beta_0 \leq \frac{2\pi}{3} b$,
and the phase transition
to the AdS-Schwarzschild black hole geometry takes place.
Then the energy of the AdS-Schwarzschild black hole 
geometry is given by
\bea
E 
= \frac{\pa I}{\pa \beta_0}
=\frac{\mbox{Vol}(S^3)3(r_+^4-r_+^2b^{2})}{16\pi G_N b^2}
=M ,
\eea
and the entropy is given by
\bea
S 
= 
\beta_0E-I=
\frac{1}{4G_N} r_+^4 \mbox{Vol}(S^3)
=\frac{A}{4 G_N} ,
\eea
where $A$ is the area of the horizon.

Actually, 
there are two black hole solutions for given
$\beta_0 < \frac{\pi}{\sqrt{2}} b$.
Here, we are interested in the
absolute minimum of the classical action.
This corresponds to choosing
the larger solution for $r_+$
for given $\beta_0$ in (\ref{b0}).
The larger $r_+$ solution is often called
big black hole, when compared with the
smaller $r_+$ solution which is often called
small black hole.
The small black hole 
has negative specific heat
and hence unstable.
It reduces to the familiar
Schwarzschild black hole in flat space
in the $b \rightarrow$ limit.
(For the big black hole, $r_+$ becomes infinite
in the $b \rightarrow$ limit,
so there is actually no 
finite flat space limit.)
We only consider the classical 
minimum of the action in our 
saddle point approximation, 
i.e. the big black hole
in the high temperature phase.

\subsection{Gravity description
of confinement/deconfinement 
in strongly coupled Yang-Mills theories}

Witten has given some gravity descriptions for
expected phenomena in strongly coupled Yang-Mills theory
\cite{Witten:1998qj,Witten:1998zw}.
Here we will recall few of them
which are most relevant for us in the
later discussions.

One of the order parameters for the deconfinement
is the action itself.
In the large $N$ expansion in
Yang-Mills theory,
the action scales like $O(N^0)$
in the confined phase and $O(N^2)$
in the deconfined phase.
In the gravity side,
the action scales like $1/G_N \sim 1/{g_s^2}$
($g_s$ is the closed string coupling)
in the black hole phase
which corresponds to the deconfined phase.
This 
matches with the Yang-Mills side
via the identification
of the 't Hooft expansion 
with the genus (loop) expansion of the closed string 
theory
(section \ref{EtH}).

\begin{figure}
\begin{center}
 \leavevmode
 \epsfxsize=60mm
 \epsfbox{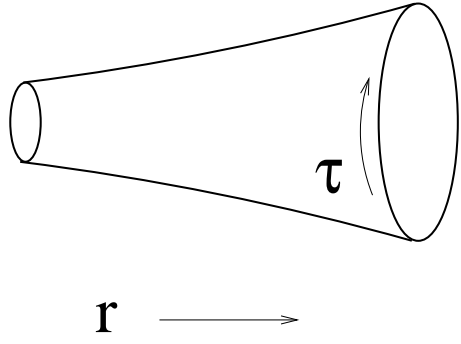}\\
\end{center}
\caption{The topology of the thermal AdS geometry.
$\tau$ and $r$ directions are depicted.}
\label{topoAdS}
\end{figure}

Another important 
order parameter is the expectation value
of the Polyakov loop,
or the Wilson loop wrapping around
the Euclidean time direction, defined by 
\bea
{\cal P} \equiv 
\frac{1}{N} \tr P e^{i \int_0^\beta d\tau A_0},
\eea
where $P$ denotes the path ordering.
We will explain more about
the Polyakov loop
in Yang-Mills theory 
in the next section.
In the bulk side in the 
classical gravity
approximation,
the expectation value of the
Wilson loop
is calculated
from the (regularized) area of 
the minimal surface
ending on the loop
\cite{Rey:1998ik,Maldacena:1998im,%
Rey:1998bq,Brandhuber:1998bs}
(see also \cite{Aharony:1999ti,Semenoff:2002kk}
for a learning):\footnote{Precisely speaking,
what is calculated in this way is
a generalization of the Wilson loop
including scalar fields \cite{Maldacena:1998im}.
However, 
in the low temperature limit,
all fields other than $A_0$ 
on $S^3$ become massive 
if dimensionally reduced and the generalized
Wilson loop reduces to the conventional one
at low energy.
On the other hand, in the high temperature
limit those fields are expected
to aquire a mass of a
scale given by the temperature,
and at distance much longer than that
the generalized
Wilson loop again reduces to the conventional one.}
\bea
\langle {\cal P} \rangle
\sim
e^{-T_{st}{\cal A}},
\eea
where ${\cal A}$ is the area of 
the minimal surface in the bulk
ending on the loop
and $T_{st}$ is the tension of 
fundamental string.
As we will explain in the next section,
in the confined phase
the expectation value of the Polyakov loops
vanishes.
In the dual gravity side,
this is explained as follows.
The thermal AdS geometry
(the AdS geometry periodically identified
in $\tau$-direction)
has a topology $S^1\times B^4$
($B^4$ stands for the four-dimensional ball), and
there is a non-contractible circle
wrapping in the thermal circle $S^1$,
see Fig.\ref{topoAdS}.
This prevents
the worldsheet with
a disk topology
to end on the thermal circle at the boundary.
This means that
the Polyakov loop expectation value is zero
at the closed string tree level, or
in the $N \rightarrow \infty$ limit.
Thus the thermal AdS geometry corresponds to
the confined phase.
\begin{figure}
\begin{center}
 \leavevmode
 \epsfxsize=65mm
 \epsfbox{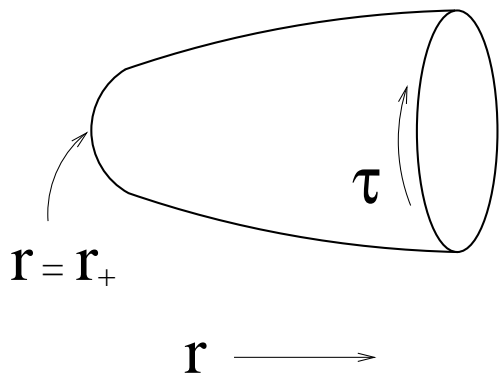}\\
\end{center}
\caption{The topology of the AdS-Schwartzchild geometry.}
\label{topoSch}
\end{figure}
On the other hand,
for the AdS-Schwarzschild black hole geometry
has a topology of  $R^2\times S^3$,
and hence there's no non-contractible circle,
see Fig.\ref{topoSch}.
In this case, a worldsheet with a disk topology
can end on the thermal circle 
on the boundary.
Then the (regularized) area of the surface gives the
Polyakov loop expectation value.
Therefore, the AdS-Schwarzschild black hole geometry
should correspond to the deconfined phase
where the Polyakov loops have expectation values.

If one compares 
the AdS-Schwarzschild black hole geometry (\ref{metAdSBH})
with the AdS geometry (\ref{metAdS}),
one notices that in the black hole geometry
the space-time beyond the horizon $r=r_+$
looks as disappeared.
It is well known that
a closed string winding around a circle
with anti-periodic boundary conditions 
for fermions
becomes tachyonic when
the circle shrinks to
the string scale \cite{Rohm:1983aq}.
It has been argued that 
the above change of the space-time geometry
is caused by condensation of
such tachyonic closed string.\footnote{%
Clarifying comment:
When the order of the phase transition
is first order, generically
(though it depends on the shape of the 
potential for the winding mode)
the phase transition
takes place before the tachyonic mode
appears. In other words,
when the phase transition
takes place,
the winding mode is still not tachyonic 
in the thermal AdS geometry.}
Analogous to the open string
tachyon condensation
to ``no-D-brane" vacuum \cite{Sen:1998sm},
the condensation of the closed string
tachyon was speculated 
to lead to the ``no-space-time" phase
and 
eliminate the space-time beyond $r\sim r_+$
\cite{KalyanaRama:1998cb,%
Barbon:2001di,Barbon:2002nw,Barbon:2004dd,%
McGreevy:2005ci,Horowitz:2006mr,Furuuchi:2006st}.
We will explore more on this point
from the Yang-Mills side 
using the AdS-CFT correspondence.
\begin{figure}
\begin{center}
 \leavevmode
 \epsfxsize=60mm
 \epsfbox{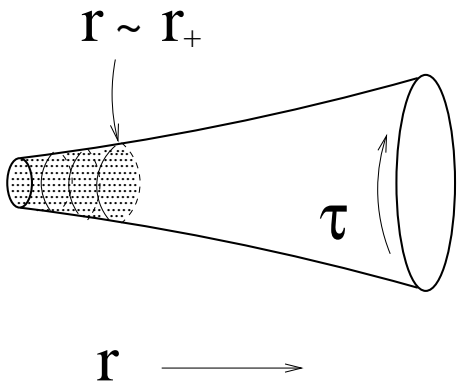}\\
\end{center}
\caption{At certain temperature,
the radius of the thermal circle reaches
string scale in the region
beyond $r \sim r_+$.
The space-time becomes
unstable due to the closed string winding
tachyon condensation,
and expected to decay to
``no-space-time" phase.
The endpoint is expected to be the
AdS-Schwarzschild geometry Fig.\ref{topoSch}.}
\label{tachyon}
\end{figure}

From the next section we will analyze
weakly coupled Yang-Mills theories.
Since the analysis in this section
is based on 
the Einstein-Hilbert action,
it is valid in the
strong 't Hooft coupling region,
according to 
the relation between the 't Hooft coupling
and the curvature radius of the
AdS (\ref{Rlambda}).
However, the existence of the 
confinement-deconfinement transition
seems robust in the Yang-Mills side,
and therefore we expect that
this kind of gravitational phase transition 
persists with appropriate string corrections,
all the way up to the weak 't Hooft coupling region.
Although the detailed form of the metric will
be modified,  we except that
general features like the difference
of topology in two geometries 
remain the same.\footnote{%
See \cite{Harmark:2006ta} for an investigation of 
a limit where both Yang-Mills side
and string theory side are weakly coupled.}

\section{Confinement-Deconfinement Transition 
at Weak Coupling}%
\label{DecW}

\subsection{Polyakov loop as an order parameter}

We will study
the gauge theory action in which
all fields are in the adjoint representation
of the $SU(N)$ gauge group.
To discuss canonical ensamble,
the Euclidean time direction is
compactified to a thermal circle with
the circumference $\beta$.
The generator of the
gauge transformation
should satisfy
\bea
 \label{pgauge}
g(\tau+\beta) = g(\tau),
\eea
in order not to change the boundary conditions
for the fields.
However, since all the fields
are in the adjoint representation,
they
transform trivially under the
center of the gauge group.
In this case 
we can consider more general
transformation
\bea
\label{ZN}
g(\tau+\beta) = g(\tau)h  ,
\eea
where $h$ is an element of 
the center of the gauge group
which can be identified with the $N$-th root of unity
$e^{2\pi i\frac{n}{N}}, \, n=1,\cdots,N-1$.

An order parameter
for the $Z_N$ symmetry
is given by the 
expectation value of 
the Polyakov loop
defined by
\bea
{\cal P} 
\equiv 
\frac{1}{N} \tr P e^{i \int_0^\beta d\tau A_0},
\eea
where $P$ denotes the path ordering
and $\tr$ is the trace in the fundamental 
representation of the gauge group.
Physically, the Polyakov loop
${\cal P}$ measures the exponential of
the free energy 
for adding an external source quark (a field in the
fundamental representation of the $SU(N)$ gauge group).
In the confined phase,
adding an external quark source
costs infinite free energy,
and thus the expectation value of the
Polyakov loop is zero.
To be more precise,
actually on a compact space like $S^3$,
due to the 
Gauss's law constraints
the expectation value of the single Polyakov loop 
is automatically zero.
However, 
the way it becomes 
zero is different in the confined phase
and the deconfined phase.
In the confined phase a
single large $N$ saddle point 
(see below)
gives zero
whereas in the deconfined phase
each saddle points 
gives non-zero value but
the sum over the
$N$ saddle points related by
the $Z_N$ symmetry
gives zero.
It is not difficult to
construct a related order parameter which probes
the phase on the compact space.
The operator $\langle | {\cal P} |^2\rangle$
which amounts to 
adding external quark and anti-quark,
can do the job.%
\footnote{See \cite{Aharony:1998qu} for more
about the $Z_N$ symmetry in AdS-CFT.}

For later purpose,
we also define a straightforward
generalizations of
the Polyakov loop,
the Wilson loops
wrapping around the thermal circle 
for $n$ times:
\bea
{\cal P}_n
\equiv 
\frac{1}{N} \tr P e^{i\int_0^{n\beta} d\tau A_0}.
\label{Pn}
\eea
We will also call these 
as Polyakov loops.

Usually, there's no phase transition
in a field theory
on a compact spatial manifold.
However, if one takes the size of the gauge group
$N$ to infinity, 
the large $N$ phase transition can take place
\cite{Gross:1980he,Wadia:1979vk,Wadia:1980cp}.

Now suppose we have
a partition function
of a theory with fields $\phi^i$
coupled to the gauge field:
\bea
Z
=
\int {\cal D}A_0
\int {\cal D}\phi^i \,
e^{-S(A_0,\phi^i)}.
\eea
To determin whether
the confined phase or
the deconfined phase
is realized,
we
integrate over
all the massive fields which\footnote{%
In the cases we will be interested in the later section,
there are no massless fields.}
to obtain 
the effective action $S_{eff}(A_0)$
for the temporal component of the 
gauge field $A_0$:
\bea
Z
=
\int {\cal D}A_0\,
e^{-S_{eff}(A_0)}.
\eea
The phase can be known by calculating the
expectation values of the
Polyakov loops like
\bea
\langle |{\cal P}|^2 \rangle
=
\int {\cal D}A_0\,
|{\cal P}|^2
e^{-S_{eff}(A_0)},
\eea
as explained above.

\subsection{Deconfinement transition in a toy model}

Now we present an explicite toy model 
example.\footnote{The explanations in this subsection
is based on \cite{Furuuchi:2003sy}.}
This toy model still
captures the essential points
of the confinement-deconfinement
phase transition
of the large $N$ gauge theories
on $S^3$ at weak coupling,
which we are interested in.
The action we will consider is
\bea
S
= 
\int_0^\beta d\tau
\,
\tr \frac{1}{2} 
(D_0\Phi^i D_0\Phi^i + \w^2 \Phi^2)
\quad (i = 1,\cdots, d).
\eea
We assume that $\w^2$ is positive (non-zero)
since that is the case if one does
similar calculations 
in Yang-Mills theories 
on $S^3$.

We can choose the
gauge in which $A_0$ is constant and diagonal:
\bea
 \label{gfix}
\frac{\pa}{\pa \tau} A_0 = 0, \quad A_{0 ab} = \delta_{ab} A_a .
\eea
However one cannot gauge away
the constant mode of $A_{0}$ when the
Euclidean time is compactified on a circle.
Recall (\ref{pgauge}).
A shift of an eigen-value $A_a$
by ${2\pi}/{\beta}$
can be canceled by the gauge transformation
of the form $g(\tau) = \mbox{diag}
(1,\cdots,1,e^{\frac{2\pi i}{\beta}\tau},1,\cdots,1)$ 
which is periodic: $g(\tau+\beta) = g(\tau)$,
and thus is a legitimate
gauge transformation.
Therefore the eigen-values
$A_a$ live on a circle
with radius $1/\beta$.

The Faddeev-Popov determinant for the first
of the gauge fixing conditions (\ref{gfix}) is
\bea
{\det}'
\left(-\frac{d}{d\tau}
 \left(
-\frac{d}{d\tau}+i(A_a-A_b)
 \right)
\right).
\eea
The prime on the det means that
the zero-mode of the time derivative
is omitted from the determinant.
The Faddeev-Popov determinant for
the diagonalization of the gauge field
is the familiar Vandermonde determinant:
\bea
\prod_{a\ne b} (A_a-A_b).
\eea
Combining them together,
we obtain
\bea
{\det}'
\left(-\frac{d}{d\tau}\right)
\det
\left(
-\frac{d}{d\tau}+i(A_a-A_b)
 \right).
\eea
Using the formula
\bea
\det
\left(
-\frac{d}{d\tau}+\w
\right)
=
2\sinh \frac{\beta \w}{2}
\eea
for the periodic boundary conditions,
we obtain
\bea
Z
=
\int \prod_{a=1}^N dA_a
\prod_{a\ne b}
\sin \frac{\beta}{2} |A_a-A_b|
\prod_{a\ne b}
\left(
\frac{1}{\sinh \left( \frac{\beta}{2} \left(\w+i(A_a-A_b)\right) \right)}
\right)^d  .
\eea
Thus in this case we obtain the following
effective action
for $A_0$:
\bea
\label{Seff}
&&S_{eff}(A_0) \nn
&&\quad =
\sum_{a\ne b} d 
\ln 
\left(
\sinh \left(\frac{\beta}{2} \left(\w+i(A_a-A_b)\right)\right)
\right)
- \ln \sin \frac{\beta}{2} |A_a-A_b| .
\eea
In the
large $N$ limit,
$S_{eff}$ is order $O(N^2)$, so we can apply
the saddle point method.
We define the
eigen-value density $\rho(\theta)$ by
\bea
\rho(\theta)
\equiv
\frac{1}{N}
\sum_{a=1}^N
\delta(\theta - \beta A_a) ,
\eea
where we can set
$-\pi \leq \theta \leq \pi$ without loss of generality.
$\rho(\theta)$ is normalized as
\bea
\int_{-\pi}^{\pi} d\theta
\rho(\theta)
=1.
\eea
Since $\rho(\theta)$ is a density,
it also satisfies
\bea
\rho(\theta) \geq 0 .
\eea
Using the eigen-value densty $\rho(\theta)$,
(\ref{Seff}) can be rewritten as
\bea
&&S_{eff} \nn
&&=
N^2
\int_{-\pi}^\pi d\theta
\int_{-\pi}^\pi d\theta'
\rho(\theta) \rho(\theta')
\left\{
d \ln 
\left( 
 \sinh 
  \left(
  \frac{\beta \w}{2} + \frac{i}{2}(\theta-\theta')
  \right)
\right)
-
\ln 
\left(
\sin \frac{1}{2} |\theta-\theta'|
\right)
\right\}.\nn
\label{Seffrho}
\eea
It is instructive to
Fourier expand the eigen-value density 
$\rho(\theta)$:
\bea
S_{eff}
=
N^2
\sum_{n=1}^\infty
\rho_n\rho_{-n}
\frac{1-d e^{-n\beta \w}}{n} ,
\label{Smodes}
\eea
where
\bea
 \label{rhon}
\rho_n
\equiv
\int_{-\pi}^\pi d\theta \rho(\theta) e^{i n\theta}
=
\frac{1}{N} \sum_{a=1}^N e^{in\beta A_a}
=
\frac{1}{N} \tr U^n .
\eea
Thus the $n$-th Fourier mode $\rho_n$
is nothing but the Polyakov loop ${\cal P}_n$
in (\ref{Pn}) which
wind around the thermal circle $n$ times.
We obtain the saddle point equation from (\ref{Seffrho}):
\bea
\label{saddleeq}
\int d\theta' \rho(\theta')\,
i\, \mbox{coth} 
\left(
\frac{\beta\w}{2}+\frac{i}{2}(\theta-\theta')
\right)
=
\int d\theta' \rho(\theta') \cot \frac{1}{2}(\theta-\theta'),
\eea
where it is understood that the principal value
around the singularity is
taken in the integral on the right hand side.

The homogeneous eigen-value density distribution
$\rho(\theta)= \frac{1}{2\pi}$
is always a solution to the
saddle point equation (\ref{saddleeq}).
Since the Polyakov loop 
expectation
values vanish in this case,
it corresponds to the
low temperature confined phase.

As we increase the temperature,
from (\ref{Smodes})
one sees that 
at $\beta=\beta_c\equiv 
\frac{\ln d}{\w}$
the $n=1$ mode becomes unstable.
This leads to  
the deconfinement phase transition.
It may be worthwhile to note
that the transition does not occur
at finite temperature in the $d=1$ case.
(This can be seen from the fact
that for $d=1$ the density of 
gauge singlet states
doesn't show the Hagedorn behavior 
\cite{Sundborg:1999ue,Polyakov:2001af,%
Aharony:2003sx,Furuuchi:2003sy}.)

One might have felt little bit 
odd that
there is a confined phase
in the weakly coupled gauge theory.
When one discusses about
the strength of the coupling,
one uses canonically normalized fields.
However,
the constant mode of $A_0$
does not have a kinetic term, 
and therefore it is always strongly coupled,
even when one takes the 't Hooft coupling small.

For further readings
for how to determine the 
eigen-value distributions
in the high temperature phase,
see \cite{Semenoff:2004bs}.
For the study of Yang-Mills theories
on $S^3$, which is 
of our actual interest, see
\cite{Aharony:2003sx,Aharony:2005bq,%
Alvarez-Gaume:2005fv}.

\section{'t Hooft-Feynman Diagrams at Finite Temperature
-- Imaginary Time Formalism}\label{EtH}

\subsection{'t Hooft expansion}

Let us consider a lagrangian of the form
\bea
{\cal L}(\Phi^i)
=
\frac{1}{g_{YM}^2}
\tr
\left\{
\pa_\mu \Phi^i \pa_\mu \Phi^i
+ V(\Phi^i)
\right\} ,
\eea
where $\Phi_{ab}^i$ are fields in
the adjoint representation of
the $SU(N)$ gauge group.\footnote{We 
will neglect the subleading
differences in ${1}/{N}$ between 
$SU(N)$ and $U(N)$ gauge group 
in the discussion here.}
\begin{figure}
\begin{center}
 \leavevmode
 \epsfxsize=80mm
 \epsfbox{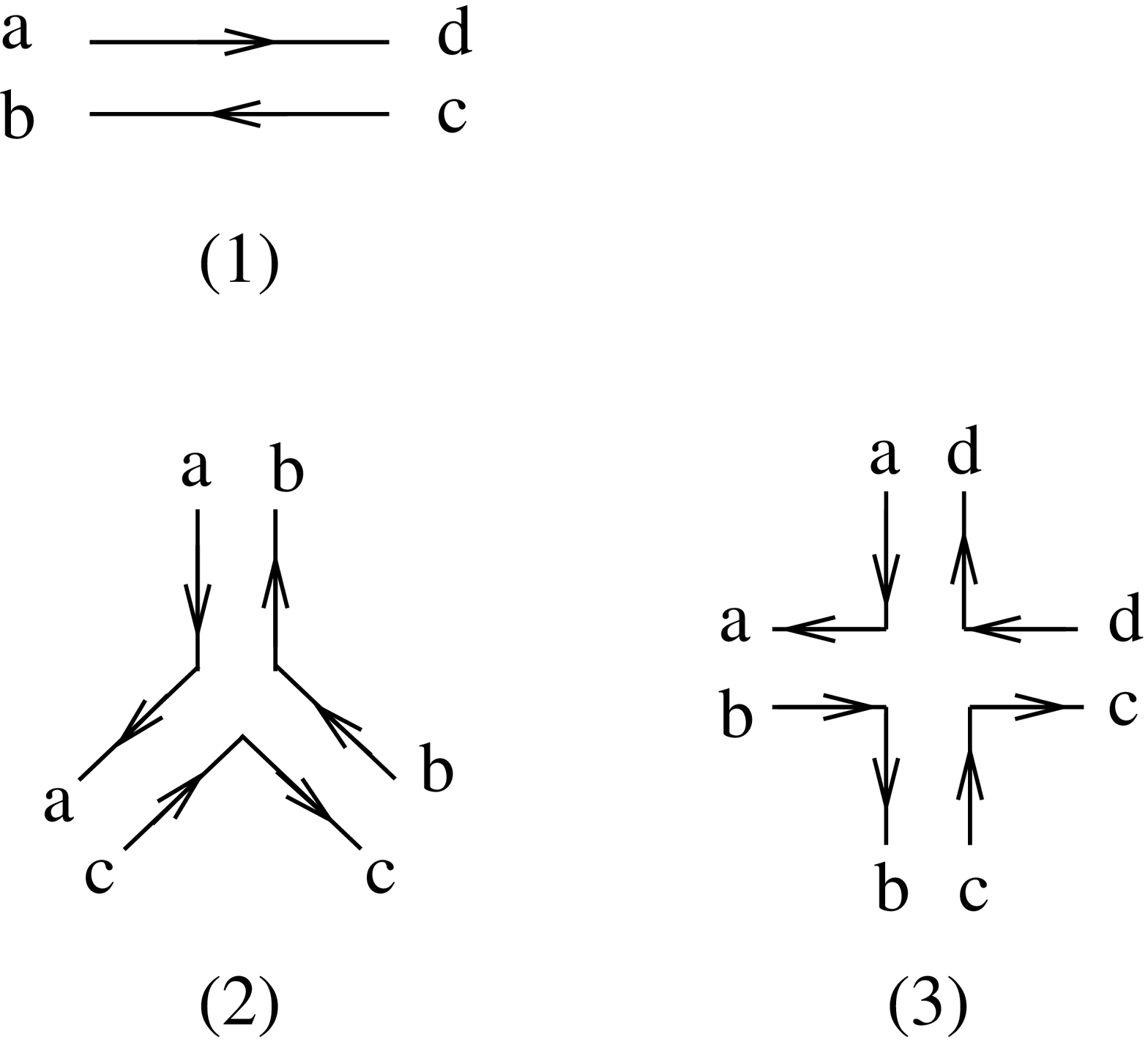}\\
\end{center}
\caption{(1) is the propagator
$\langle \Phi_{ab}\Phi_{cd} \rangle$.
(2) and (3) schematically denote the
interaction vertices,
just showing how 
the flow of the matrix indices
are drawn.}
\label{tHooft}
\end{figure}
't Hooft noticed that if one takes
the large $N$ limit
$N \rightarrow \infty$ together with
$g_{YM} \rightarrow 0$\footnote{%
For the ${\cal N}=4$ super Yang-Mills theory,
the coupling does not run.
In general the coupling runs and this 
expansion
becomes formal.
However on a compact manifold 
$S^3$, if we set
the size of its radius $R$ to be
much smaller than the scale
$\Lambda_{QCD}$ dynamically generated
from the asymptotically free
Yang-Mills theory, i.e.
$R\Lambda_{QCD} \ll 1$,
the 't Hooft limit makes sense.} with 
the 't Hooft coupling 
$\lambda \equiv g_{YM}^2 N$ 
held fixed, 
Feynman diagrams of such theories are 
organized into simplicial decompositions
of Riemann surfaces \cite{'tHooft:1973jz}.
\begin{figure}
\begin{center}
 \leavevmode
 \epsfxsize=85mm
 \epsfbox{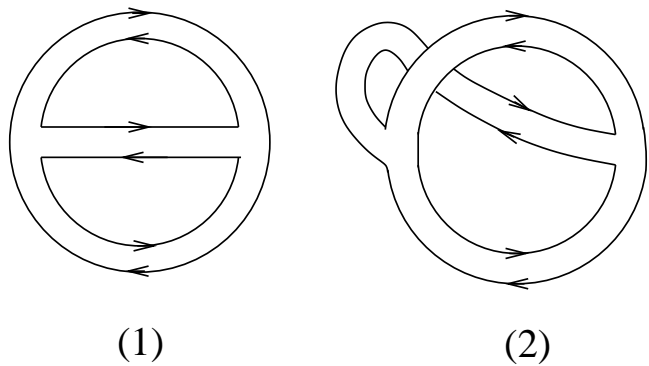}\\
\end{center}
\caption{Examples
of 't Hooft-Feynman diagrams.
(1) can be drawn on a sphere (genus zero surface) 
whereas
(2) can be drawn on a torus (genus one surface) (try).}
\label{tHdiagrams}
\end{figure}
%
To see this,
we explicitely follow the flow of 
the matrix indices
using the 't Hooft's double line notation
depicted in Fig.\ref{tHooft}.
Then, for a given Feynman diagram,
each interaction vertex contributes
by a factor of ${1}/{g_{YM}^2}={N}/{\lambda}$,
and each propagator contributes by a factor
of ${\lambda}/{N}$,
and each index loop contributes by a factor
$N$ since it contains a sum over $N$ gauge indices.
Therefore, a diagram with
$V$ vertices (vertices of the simplicial decomposition of the surface on which the diagram can be drawn), 
$E$ propagators (edges of the simplicial decomposition)
$F$ index loops (faces of the simplicial decomposition)
contributes with a factor
\bea
N^{V-E+F} \lambda^{E-V}
=
N^\chi \lambda^{E-V},
\eea
where $\chi \equiv V-E+F$ is the Euler character
of the surface on which the diagram can be drawn.
It is $\chi = 2-2g$ for closed oriented surface
with genus $g$ (the number of handles).\footnote{Here we are discussing connected diagrams. For disconnected
diagrams we have such contributions 
from each connected component.}
Therefore, if we identify 
the Riemann surfaces with closed string worldsheets,
the closed string coupling is read as
\bea
g_s \propto \frac{1}{N}.
\eea

\subsection{'t Hooft expansion at finite temperature
and Polyakov loops}

Now we study the perturbative expansion
around the saddle points
of $A_0$ discussed in section \ref{DecW}.
The covariant derivative for
adjoint fields is given by
\bea
D_0 \Phi(\tau)
=
\pa_0 \Phi(\tau) - i [A_0,\Phi(\tau)].
\eea
We have suppressed the spatial dependence
cause it it not essential in the following
discussions.
Bosonic fields obey the periodic boundary condition
\bea
\Phi(\tau+\beta)= \Phi(\tau).
\eea
(One can make similar arguments for fermionic fields.)
It is convenient to make a
field redefinition
\bea
\tPhi(\tau) = e^{-i A_0 \tau} \Phi(\tau) e^{i A_0 \tau} .
\eea
Then the covariant derivative is transformed
into normal derivative:
\bea
\pa_0 \tPhi(\tau) 
= e^{-i A_0 \tau} D_0 \Phi(\tau) e^{i A_0 \tau}
\eea
And the 
boundary condition for $\tPhi(\tau) $ becomes
\bea
\tPhi(\tau+\beta)= U ^{-1}\tPhi(\tau)U ,
\eea
where
\bea
U \equiv e^{i \beta A_0} .
\eea
The propagator
on $S^1\times S^3$
can be obtained 
from the propagator on ${R}\times S^3$
by the method of images:
\bea
\langle
\tPhi_{ab}(\tau)\tPhi_{cd}(0)
\rangle_{S^1\times S^3}
=
\sum_{n=-\infty}^{\infty}
\langle
(U^{-n} \tPhi(\tau+n\beta)U^{n})_{ab} \tPhi_{cd}(0)
\rangle_{{R}\times S^3}.
\label{prop}
\eea
We are interested in
calculating correlation functions of
local gauge invariant single trace operators.
In the AdS-CFT correspondence,
these correlation functions
correspond to closed string amplitudes.
In terms of the propagator on ${R}\times S^3$,
each propagator, which corresponds to
an edge of a 't Hooft-Feynman diagram,
has sum over images appearing in the
righthand side of (\ref{prop}).
This sum in the 't Hooft-Feynman diagram can be
reorganized into
the sum over index loops, interaction vertices
and cycles on handles
through the relation
\bea
E = F + V - (2-2g) =
(F-1)+(V-1)+2g.
\eea
This can be interpreted as 
sum over $F-1$ loops of the Feynman diagram, 
$V-1$ independent positions of the 
images of the interaction vertices
(only the relative positions matter),
and two cycles associated with each handle.
Then for a given Feynman diagram,
we have the following overall factor 
from the index loops:
\bea
\langle
\tr U^{w_1} \tr U^{w_2} \cdots \tr U^{w_{F-1}} 
\rangle_{A_0}
\label{Uw}
\eea
where
\bea
\langle
F(A_0)
\rangle_{A_0}
\equiv
\int dA_0\, F(A_0) e^{-S_{eff}(A_0)} .
\label{<>A}
\eea
Here $w_{f}$ is the winding number
associated with the $f$-th loop
(Fig.\ref{winding}):
\begin{figure}
\begin{center}
 \leavevmode
 \epsfxsize=70mm
 \epsfbox{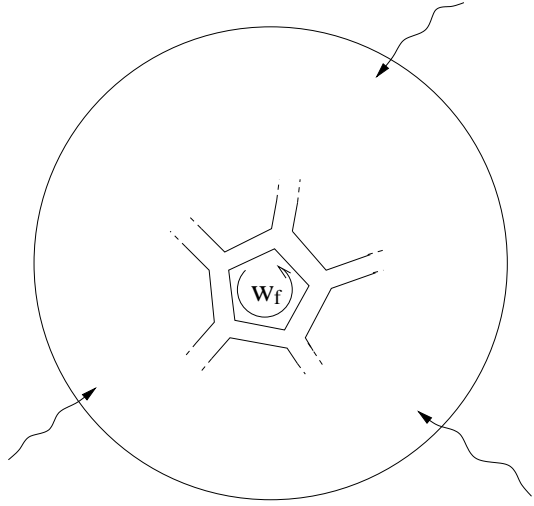}\\
\end{center}
\caption{The winding number $w_f$ is associated with
the $f$-th index loop, or the face of the simplicial
decomposition.}
\label{winding}
\end{figure}
\bea
w_{f} 
\equiv
\sum_{\mbox{\tiny Edges on $f$-th loop}} n_l ,
\eea
where $n_l$ is the number
associated with the 
$l$-th edge(=propagator) which is
on the $f$-th loop,
labeling the images
in the righthand side of (\ref{prop}).
$w_{f}$ counts how many times
the $f$-th index loop wind around 
the thermal circle.
From (\ref{Uw}),
we observe that 
{\em the expectation value
of the Polyakov loop winding $w$ times
around the thirmal circle, introduced in
(\ref{Pn}),
gives
the weight to the $w$ time-winding of a 
loop in the Feynman diagrams}.
Note that this is nothing but 
the $w$-th Fourier
mode of the eigen-states density
$\rho(\theta)$, 
(\ref{rhon}) in section \ref{DecW}.\footnote{%
We thank T. Matsuo and S. Wadia for discussions
on this point.}

So far, we were discussing
$1/N$ expansion but hasn't taken
the $N\rightarrow \infty$ limit.
The eq.(\ref{Uw})
is applicable for finite $N$ \cite{Brigante:2005bq}.
Now, let us consider the 
$N\rightarrow \infty$ planer limit.
Then, in the confined phase
all the expectation value of the
Polyakov loops vanish
and only diagrams
with zero-winding numbers
on their index loops
contribute.
This means that one needs
to sum over images only for the 
interaction vertices and
the external single trace operators.
However, the path integration over
the position on $S^1$ together
with the sum over images
recovers the path integral
over entire ${R}$ because
\bea
\sum_{n=-\infty}^{\infty}
\int_0^\beta d\tau f(\tau+n\beta)
=
\int_{-\infty}^{\infty} d\tau f(\tau)
\eea
for any function $f$.
That means, 
{\em the path integral for the internal
loops
is the same 
as that of the uncompactified theory},
the theory on ${R}\times S^3$.
Thus in terms of the correlation functions
on $R\times S^3$, the correlation functions
on $S^1\times S^3$ in the confined phase
is expressed as
\bea
&&
\langle 
{\cal O}_1(\tau_1)
{\cal O}_2(\tau_2)
\cdots
{\cal O}_n(\tau_n)
\rangle_{S^1\times S^3} \nn
&=&
\sum_{m_1,\cdots, m_{n-1}=-\infty}^{\infty}
\langle 
{\cal O}_1(\tau_1+m_1\beta)
{\cal O}_2(\tau_2+m_2\beta)
\cdots
{\cal O}_n(\tau_n)
\rangle_{R\times S^3},
\label{composits}
\eea
where ${\cal O}_i$'s are local gauge invariant
single trace operators.
As we have mentioned above,
there are sum over images for 
the external operators but one.
Note that for composite operators ${\cal O}_i$,
(\ref{composits}) is {\em not} 
a straightforward consequence
that the time direction is compactified,
but the winding modes associated with
the index loops must be suppressed.
As we have seen,
the essential reason for this suppression
was that we were in the confined phase.
This mechanism 
that the result of the compactified theory
can be obtained from the un-compactified theory
(or vise versa),
or in other words the correlation functions
do not essentially depend on the
radius of the compactification radius, 
is known as the large $N$ reduction
\cite{Eguchi:1982nm}.
Its role in the context of AdS-CFT was 
first noticed in \cite{Furuuchi:2005qm}
and was further investigated
in \cite{Furuuchi:2005eu}
(see also \cite{Brigante:2005bq}).
In the bulk side, this means that
the geometry corresponding to
the confined phase is
the simple periodic identification of the
geometry of the zero-temperature theory,
i.e. AdS$_5$ for the ${\cal N}=4$ super Yang-Mills theory:
If one identifies ${\cal O}_i$ with the closed string
field in the bulk, (\ref{composits}) is 
nothing but the
method of images {\em in the bulk}.
(How to read off
the bulk geometry from the gauge theory
't Hooft-Feynman diagrams
is far from being completely understood,
much remains to be investigated.
However,
see \cite{Gopakumar:2003ns,Gopakumar:2004qb}
for suggestive
examples of how at zero-temperature
the bulk geometry seems to emerge 
from the gauge theory
't Hooft-Feynman diagrams.)
In section \ref{HP} we obtained the same
result in the gravity side, 
which correspond to the
strong 't Hooft coupling region.
Here, instead, we started from the perturbative
Yang-Mills side and read off the geometry.

In the deconfined phase,
the expectation value of 
the Polyakov loops
has vacuum expectation value
and from (\ref{Uw})
we observe that
the winding modes 
associated with 
the index loops
contribute.
This statement is 
valid for finite $N$.
In the $N \rightarrow \infty$ limit,
we can apply the saddle point approximation,
for example, 
as we have seen in
section \ref{DecW}.
In the simple toy model
there, the 
deconfinement transition temperature
was the point
where the first 
Fourier mode of the eigen-value density,
or the winding number one Polyakov loop,
becomes unstable
See eq.(\ref{Smodes}) and the explanations there.
In order to obtain closed string worldsheets
from the 't Hooft-Feynman diagrams,
one must somehow ``glue" the face of the
't Hooft-Feynman diagrams.\footnote{%
See \cite{Gopakumar:2004qb,Gopakumar:2005fx} 
for a proposal
for the precise prescriptions for this
glueing.}
However,
the non-zero winding number on a face
is an obstruction for such glueing.
This makes closed string interpretation
difficult
in the (trivial periodic identification of the)
zero-temperature bulk geometry.
From the closed string
point of view,
these winding modes associated with 
the index loops
corresponds to 
vortices on the worldsheet.\footnote{%
The worldsheet theory should have
a coordinate field $\tau(\sigma^0,\sigma^1)$
corresponding to the $\tau$ direction in the target space,
where $\sigma$'s
are worldsheet coordinates.
The winding in the target space 
$\tau$ direction means
$\tau(\sigma^0,\sigma^1+2\pi)
=  \tau(\sigma^0,\sigma^1) + \w \beta$,
where $\w$ is the winding number. 
Let us change the worldsheet coordinates to
$z=\exp(\sigma^0+ i\sigma^1)$. 
Then, $\w$ is
the winding number
counting how many times the field $\tau(z)$
wrap around the target space circle
as we go around $z=0$:
$z \rightarrow z e^{2\pi i}$.
Such configuration is called 
vortex with vorticity $\w$.
$\tau(z)$ should corresponds to the
$\tau$ coordinate of a point on the random surface
given by the 't Hooft-Feynman diagram.}
Therefore, we have seen that
the deconfinement corresponds to the
vortex condensation on the worldsheet.\footnote{%
This type of phase transitions 
in two dimensions
caused
by the vortex condensation 
are called Berezinsky-Kosterlitz-Thouless transition 
\cite{Berezinsky:1970pd,Berezinsky:1972aa,%
Kosterlitz:1973xp}.
See for example 
the insightful book \cite{Polyakov:1987ez}
for explanations,
from which
the contents of \cite{Berezinsky:1970pd}
can partially be known.}
Notice that
the above explanation is based on the
use of the propagator
on ${R}\times S^3$,
i.e.
the righthand side of (\ref{prop}).
In the bulk, this corresponds to the expansion around
(a simple periodic identification of)
the zero-temperature geometry.
If we want a closed string worldsheet
interpretation,
this is not an appropriate background
in the deconfined phase.
The target space geometry must be
deformed by the vortex/winding mode condensation
on the worldsheet.
The geometry after the vortex/winding mode condensation
should be probed by using the
expression of the propagator
in the lefthand side
of (\ref{prop})
(i.e. the form {\em after} summed over).
Since there is no sum over images
in the lefthand side of (\ref{prop}),
this should correspond
to the bulk geometry
without
a non-contractible circle,
see Fig.\ref{topoSch}.
And since the there's no
winding mode if we use the lefthand side
of (\ref{prop}), it should in principle
be possible to
``glue" the faces of the 't Hooft-Feynman diagram.\footnote{%
Although we do not yet have a prescription for glueing
as precise as that in the zero temperature
case of \cite{Gopakumar:2004qb,Gopakumar:2005fx} in this case.}
Thus in the 't Hooft-Feynman diagrams,
the difference of the bulk topology
corresponds
to whether or not
it is appropriate to use
the propagator on $S^1 \times S^3$
(lefthand side of (\ref{prop})) or 
the propagator on ${R}\times S^3$
(righthand side of (\ref{prop})).
Hence the 't Hooft-Feynman diagrams
at least can probe the difference of
the topology of the bulk geometry.
Compare the expectation from the
gravity/string side in section \ref{HP}.

\begin{figure}
\begin{center}
 \leavevmode
 \epsfxsize=120mm
 \epsfbox{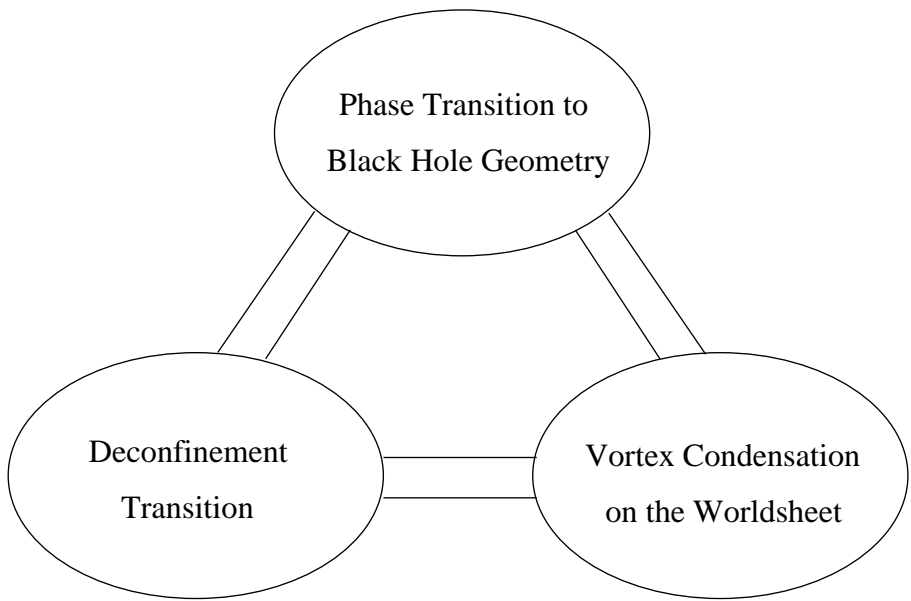}\\
\end{center}
\caption{The suggested equivalence of
the three phase transitions.}
\label{triangle}
\end{figure}

Thus, from the analysis of the
't Hooft-Feynman diagrams at finite temperature,
we arrived at the picture that
{\em the Hawking-Page transition to the
black hole geometry,
the deconfinement transition in gauge theories,
and the vortex condensation
on the closed string worldsheets
are all equivalent} (Fig.\ref{triangle}).
This is the first main message of this lecture note.
%
Of course,
this equivalence should be made
more precise by
making the AdS-CFT conjecture 
more concrete.

It has been expected that
the Hagedorn transition in string theory
indicates the appearance of
more fundamental degrees of freedom,
like the deconfinement transition in the gauge theory
\cite{Atick:1988si}.
The origin of the Hagedorn transition
was identified with 
the vortex condensation on string worldsheets
in \cite{Sathiapalan:1986db,Kogan:1987jd}.
It is intriguing that 
the above picture we have arrived 
tells that
the Hagedorn transition {\em is} 
the deconfinement transition, 
the equivalence being given via the AdS-CFT duality.
From this point of view,
it is quite curious that
we still seem to have
closed string description 
even after the deconfinement,
although 
gauge non-singlet states 
which are not identified as close string
excitations also begin to play a role.

It may be worthwhile to point out
that the order of the phase transition
is model dependant.
But the analysis of the
't Hooft-Feynman diagram
does not get modified much.
This is because
the phase is
determined by the 
effective action of the
eigen-values studied in the previous section,
and the 't Hooft-Feynman diagrams just probe it.
Most Yang-Mills theories on $S^3$ 
at weak coupling seem to
give rise to the first order deconfinement
phase transition
\cite{Aharony:2003sx,Aharony:2005bq}.
That
means the phase transition generically
could occur before the
tachyonic mode appears.
But whatever the order of the phase transition is,
the deconfined phase corresponds to the
vortex-condensate phase on the worldsheet.

The analysis of the 't Hooft-Feynman diagrams
in this section is closely related to
those in
\cite{Kazakov:2000pm,Alexandrov:2001cm,Alexandrov:2003ut}
studied in the context of a matrix model for 
the two-dimensional black hole.
(See \cite{Gross:1990md,Gross:1990ub,Boulatov:1991xz} 
for earlier studies on vortex condensation
in matrix models.)
If we regard the AdS-CFT correspondence
as a particular example
of more general 
open-closed string duality,
our AdS$_5$-CFT$_4$ model and
the two-dimensional model
may be looked from a unified view.\footnote{%
This view was already there in \cite{Kazakov:2000pm}.
That the 
matrix model for the two-dimensional black hole
can be practically analyzed
in the way parallel to the
AdS-CFT case 
was stressed in \cite{Suyama:2004vk},
where the expectation value of the
Polyakov loop was calculated both from
the matrix model and from the closed string theory.}
The analysis of
the 't Hooft-Feynman diagram
in  AdS-CFT at finite temperature
was studied in \cite{Furuuchi:2005qm}
in the context of the
Gopakumar's program 
\cite{Gopakumar:2003ns,Gopakumar:2004qb,%
Gopakumar:2005fx,Gopakumar:2004ys,%
Aharony:2006th,David:2006qc}.
See also \cite{Basu:2005pj,Brigante:2005bq}
for related discussions.
Some explanations in this section
follow the reference \cite{Brigante:2005bq}.
Although there is an advantage
in seeing the two-dimensional model
and AdS-CFT from the unified view-point,
one should also be aware of 
some technical differences.
In the AdS$_5$-CFT$_4$ case,
the effective action exhibits
phase transition by just 
varying the temperature,
whereas
in the two-dimensional model
one needs to deform the matrix model itself
by adding a potential 
for the eigen-values
of the gauge field
(change of the measure 
$[d\Omega]\rightarrow[d\Omega]_\lambda$
in \cite{Kazakov:2000pm}),
in order to obtain the deconfined phase
which should 
correspond to the black hole geometry.\footnote{%
In the case of AdS$_5$-CFT$_4$,
the change of the temperature in the same theory
changes the effective matrix model of $A_0$.}
Recall that in section \ref{DecW}
we needed $d\geq 2$
to have
phase transition at finite transition temperature.
This indicates it is not straightforward
to obtain $S_{eff}$
which exhibits the deconfinement transition
at finite temperature by
integrating out a field in 
matrix models for
two-dimensional target space.
Also at this moment 
the holography seems to be best understood 
in asymptotically
AdS spaces.
On the otherhand, 
the closed string
worldsheet theory seems to be 
better understood in
the two-dimensional model \cite{Kazakov:2000pm,Hori:2001ax}.
Finally,  
we may encounter more
qualitative differences
between open-closed duality based on
stable D-branes (AdS$_5$ case) 
and unstable D-branes 
(matrix models for two-dimensional strings)
if we go into more detailed examinations.

The analysis of 
the previous section and this section
clearly shows that
it is of essential importance,
particularly at finite temperature,
that the world-volume theory
on D-branes is not just a 
matrix field theory
but a gauge theory.
This point was not appreciated much
in the old studies of the matrix models,
but began to be realized after \cite{McGreevy:2003kb}.

\section{'t Hooft-Feynman Diagrams at Finite Temperature
-- Real Time Formalism}\label{LtH}

\subsection{Motivation}
In the previous sections,
we have studied 
thermodynamics in 
the Euclidean time formalism
(the imaginary time formalism).
As we have seen in section \ref{HP},
the Euclidean geometry 
only covers the region outside 
the black hole horizon.
However, the real problems
about black holes,
such as
the information loss paradox
and 
curvature singularities,
arise from inside the black hole.
Hence it is important 
to develop a method 
which can treat the
Lorentzian time signature case.
In the boundary field theory side,
this should correspond to 
a field theory at finite temperature
in the Lorentzian signature.
The formulation of a field theory
at finite temperature 
with the Lorentzian time signature 
is generally called the
real time formalism.
We will explain two such formalisms below:
One is the thermo-field dynamics
and the other is the closed-time-path formalism.%

\begin{figure}
\begin{center}
 \leavevmode
 \epsfxsize=100mm
 \epsfbox{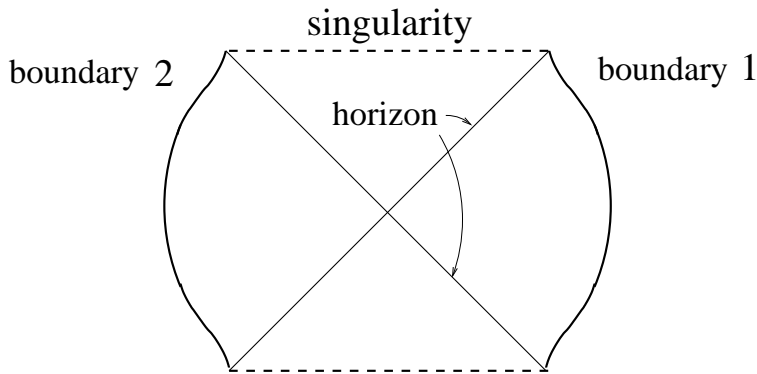}\\
\end{center}
\caption{The  Carter-Penrose diagram
of the maximally extended
AdS-Schwarzschild black hole geometry.
Spherical directions are suppressed in the figure.
Besides the usual boundary of the AdS space
at spatial infinity (boundary 1, see also Fig.\ref{AdS}),
there is a second boundary behind the horizon (boundary 2).
For more detail, see e.g. \cite{Fidkowski:2003nf}.}
\label{AdSBH}
\end{figure}

In the real time formulation of
finite temperature field theories
\cite{%
Takahasi:1974zn,Umezawa:1982nv,Semenoff:1982ev,%
Niemi:1983nf},%
\footnote{%
For further learning of
the real time formulation of
field theories at finite temperature,
see e.g.
\cite{Landsman:1986uw,LeBallac:1996bm}
for general aspects,
\cite{Umezawa:1993yq}
for insightful exploration of the 
thermo-field dynamics,
\cite{Ojima:1981ma} 
for BRS quantization of gauge theories.}
it seems almost necessary
to introduce
an additional set of fields besides
the original ones
(those in the zero-temperature theory).
Each of the newly introduced field
is associated with a field in the original theory.
(In this note we will refer to the newly introduced
fields as type-2 fields, as opposed
to the original fields
which we will call type-1 fields.)
On the other hand, the
Carter-Penrose diagram
of the maximally extended 
AdS-Schwarzschild black hole geometry
has a boundary behind the horizon
(Fig.\ref{AdSBH} boundary 2),
in addition to the usual
boundary of the AdS space at spatial
infinity outside the horizon
(Fig.\ref{AdSBH} boundary 1).
\begin{figure}
\begin{center}
 \leavevmode
 \epsfxsize=30mm
 \epsfbox{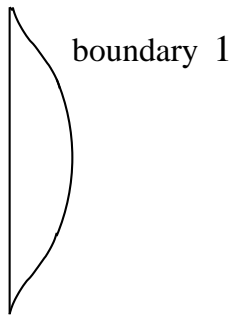}\\
\end{center}
\caption{The Carter-Penrose diagram
of the AdS geometry.}
\label{AdS}
\end{figure}
In the context of the AdS-CFT correspondence,
the
type-2 fields in the boundary CFT
was identified as 
the degrees of freedom
behind the black hole horizon 
\cite{Balasubramanian:1998de,Horowitz:1998xk,%
Maldacena:2001kr}.\footnote{%
See also \cite{Herzog:2002pc} for an approach to
describe the path integral formulation
of the real time formalism
from the gravity side in black hole phase.}
This gives a nice explanation
for the necessity of the introduction
of type-2 fields
in the real time formulation 
of the finite temperature field theory.

However, 
as we have seen in section \ref{HP},
below the Hawking-Page transition temperature
the thermodynamically preferred configuration
in canonical ensemble is
the thermal AdS without a black hole.
Since it is a finite temperature system,
the dual gauge theory
should still be described by
the real time formulation for the finite temperature.
Thus one must conclude that
below the Hawking-Page transition temperature,
the type-2 fields in the dual gauge theory
cannot correspond to the
degrees of freedom behind the black hole horizon,
since there is no black hole at all.
But then, what is the dual bulk description
for the type-2 fields in this case?
Since the AdS geometry corresponds to
the confined phase in gauge theory side,
the confinement should change
the role of the type-2 fields in the bulk.
We will show that this is indeed the case:
In the confined phase, the role of type-2 fields
in the gauge theory
is to make up {\em bulk} type-2 fields of
closed string field theory
on AdS at finite temperature
in the real time formalism.
In this section,
we will explain how this is described
from the Yang-Mills theory.
Since the bulk is also
at finite temperature,
it is natural that
the bulk theory
also has the type-2 fields of its own.
The discussions will be
in the leading order in the $1/N$ expansion,
which corresponds to
the classical theory in the bulk.
Since the Carter-Penrose diagram
is based on classical gravity,
this is sufficient
for explaining
why we should not obtain the region
behind the horizon in this case.

\subsection{Thermo-field dynamics}\label{ssecHH}

The quantities we are interested
in field theories at finite temperature
are expectation values of operators
in the canonical ensamble:
\bea
\label{Cano}
\Tr \left\{ e^{-\beta \hat{H}} 
{\cal O}_1(t_1) {\cal O}_2(t_2) \cdots {\cal O}_n(t_n) 
\right\}.
\eea
Since the Hisenberg 
operators ${\cal O}_i(t_i)$ depend on time,
the imaginary time formalism
is not applicable.\footnote{%
The relation between the real time correlation
functions and those 
analytically continued from the imaginary time
formalism are not simple in general.
See e.g. \cite{Baier:1993yh}.}
Thermo-Field Dynamics (TFD)
\cite{Takahasi:1974zn}
is a formalism in which
(\ref{Cano}) is calculated
as a kind of ``vacuum" expectation value.
For that purpose, 
one first double the
Hilbert space by preparing a copy of the
original one.
We will label the original operators and states
with suffix 1, and their copies with suffix 2.
Then we introduce
``thermal vacuum" at inverse temperature $\beta$ by
\bea
\label{vacb}
|0,\beta \rangle
\equiv
\sum_{n}
e^{-\frac{\beta E_n}{2}}
|E_n \rangle_1 |E_n \rangle_2 ,
\eea
where
$|E_n \rangle$ is the energy eigen-state with
the eigenvalue $E_n$.
The thermal vacuum is invariant
under time translation with respect to
the Hamiltonian $H_{TFD}$ defined by
\bea
\hat{H}_{TFD}=\hat{H}_1-\hat{H}_2 ,
\eea
where $\hat{H}_1$ is the original Hamiltonian,
and $\hat{H}_2$ is the copy acting on 
the copy Hilbert space.
It is easy to see that
by the use of the thermal vacuum,
(\ref{Cano}) can be written in the form of a
vacuum expectation value:
\bea
\langle 0,\beta |  {\cal O}_1 \cdots |0,\beta \rangle
=
\sum_{n} e^{-\beta E_n}
{}_1\langle E_n |  {\cal O}_1 \cdots  |E_n \rangle_1
=
\Tr_1 \left\{ e^{-\beta \hat{H}_1}
 {\cal O}_1 \cdots \right\}.
\eea
In the above operators are all original ones
which we have given the suffix 1.
Now let us consider
applying this formalism
to the large $N$ gauge theories,
in the context of the AdS-CFT correspondence.
In the confined phase,
the energy eigen-states 
are given by gauge singlets.
If we identify those states constructed
by acting
local gauge invariant single trace operators
on the vacuum
with states in the closed string theory following the 
standard AdS-CFT dictionary (see \cite{Banks:1998dd}),
we should obtain the
thermo-field dynamics 
for the closed string field theory.
As long as we are using the same 
energy eigen-states
as those of zero-temperature case,
we should also 
obtain the same bulk geometry
on a time-slice.
However, in this explanation it is not obvious
whether it is appropriate to keep on using the
zero-temperature energy eigen-states,
i.e.
why corrections by the effect of the temperature
can be neglected.
In the bulk, this corresponds to whether
we obtain the AdS geometry 
as an appropriate background or not
at finite temperature.
This question is basically solved by 
the large $N$ reduction
explained in the previous section.
But
we should look for the real time
version of the large $N$ reduction.

We will investigate this using the
path integral formulation
of the gauge theory at finite temperature
in the real time formalism.
Since the covariant path integral approach
was the most efficient way
to calculate closed string amplitudes
in critical string theories,
it would also be useful to
investigate this method in our case.
This also allows us to see
how the expectation value of Polyakov loops
we studied in section \ref{DecW}
are reflected on the closed string worldsheets.

We will mainly describe 
the confined phase case below, but
the difference from the deconfined phase
will be mentioned.
Our ultimate objective
will be to understand how
black holes are described in the
dual Yang-Mills theory.
But in order to understand the black hole,
it is important to understand first
how the situation
without a black hole is described
in Yang-Mills theory.
Once this is understood, then
we can study the difference between that
and the Yang-Mills 
description of the
black hole geometry.

\subsection{Path integral method in
the real time formalism}\label{PIr}

In this section, we will review
the path integral method
in the real time formulation of finite temperature
field theories \cite{Niemi:1983nf}.
This path integral formulation
of field theory at finite temperature
in the real time
can be regarded as a
special case of the
Schwinger-Keldysh closed-time-path formalism
\cite{Schwinger:1960qe,Keldysh:1964ud}
which was formulated with an attempt
to describe non-equilibrium systems.
We would like to apply this formalism
to the large $N$ gauge theories.
A concrete example
in mind
is the ${\cal N}=4$ super Yang-Mills
theory with $SU(N)$ gauge group on $S^3$
in the 't Hooft limit,
but we will consider
a simpler model which captures the essential point.
%
We take a real
scalar field $\Phi_{ab}(t)$
in the adjoint representation
of $SU(N)$ as an example.
Here, $a,b$ are $SU(N)$ gauge indices.
It is straightforward to include several
scalar fields,
fermions or dynamical gauge fields.
Since the dependence on the spatial $S^3$
direction is not essential in the following
argument, we can first consider
a quantum mechanical model
obtained from the dimensional reduction.
It is easy to incorporate the dependence
on the $S^3$ directions.

\begin{figure}
\begin{center}
 \leavevmode
 \epsfxsize=100mm
 \epsfbox{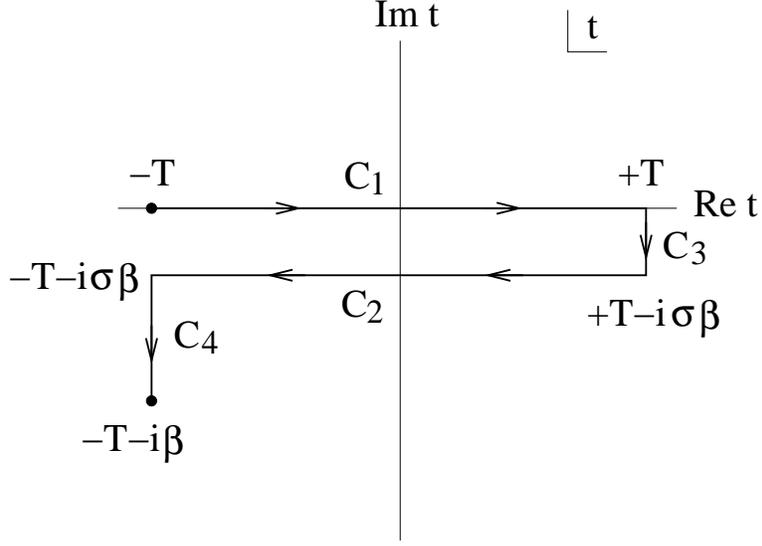}\\
\end{center}
\caption{The contour $C$}
\label{contour}
\end{figure}

The quantities we will be interested 
are
thermal Green's functions
$G_{\beta}(t_1,\cdots,t_n)$
of the time-ordered products
of operators
$\hPhi(t)$ in Heisenberg picture:
\bea
 \label{TG}
G_{\beta}(t_1,\cdots,t_n)
=
\frac{1}{\Tr\, e^{-\beta \hH}}
\Tr
\left\{ e^{-\beta \hH}
T[\hPhi(t_1) \cdots \hPhi(t_n)]
\right\},
\eea
where ``$\Tr$" is the trace over
{\em physical}
states
satisfying the Gauss' law constraints:
\bea
 \label{phys}
\hat{\rho}_{ab}\, |phys \rangle = 0.
\eea
Here
$\hat{\rho}_{ab} =
i :([\hPhi,\hPi_\Phi])_{ab}:$ 
is the generator of
the gauge transformation, where
$\hPi_{\Phi ab}$ is the conjugate momentum
of $\hPhi_{ab}$ and $:\,\,\, :$ denotes
the normal ordering.
$T[\cdots]$ above denotes the time ordering
and $\beta$ is the inverse temperature.
We will study the Hamiltonian $\hH$ given by
\bea
 \hH =
\tr
\left\{
\frac{g^2}{2}\hPi_{\Phi}\hPi_{\Phi}
+ \frac{\w^2}{2g^2} \hPhi^2 + \frac{1}{g^2} V[\hPhi]
\right\},
\eea
where
$g$ is the gauge coupling constant,
``$\tr$" is a trace over the $SU(N)$
gauge group indices and
$V[\hPhi]$ is a potential term.
The mass $\w$ is proportional to 
the inverse radius of $S^3$ in the case when
the quantum mechanics
is obtained from the compactification
of four dimensional conformal field theory 
on $S^3$.
In order to evaluate
the thermal Green's functions
by the path integral method,
we should extend the support of the field variables
to the whole complex $t$-plane:
\bea
\hPhi(t) = e^{i \hH t} \hPhi(0) e^{- i \hH t}.
\eea
We would like to obtain a
functional representation for
the generating functional $Z[J]$ such that:
\bea
G_{\beta}(t_1,\cdots,t_n)
=
\frac{1}{Z(0)}
\frac{1}{i^n}
\frac{\delta}{\delta J(t_n)} \cdots
\frac{\delta}{\delta J(t_1)}
Z[J] \Biggr{|}_{J=0}  .
\eea
The functional
\bea
\Tr
\left\{
e^{-\beta \hH}
T[e^{i \int_{-T}^{T} dt J(t)\hPhi(t)}]
\right\}
\eea
has this property when $-T < t_i < T$.
But in order to calculate the thermal
Green's function perturbatively,
we should extend the $t$ integration
to a contour $C$ on the complex plane:
\bea
\label{ZJ2}
Z[J]
=
\Tr
\left\{ 
e^{-\beta \hH}
T_C[e^{i \int_C dt J(t)\hPhi(t)}]
\right\}
\eea
where the contour $C$ is depicted in
Fig.\ref{contour}.\footnote{%
There is some arbitrariness in the
choice of the contour. Is should
be everywhere non-increasing 
in the imaginary direction
in order for the 
expression to be convergent 
in the $T \rightarrow \infty$ limit.}
$T_C[\cdots]$ denotes time-ordering
along the contour $C$.
We insert the identity
$1 = e^{-i\hat{H}T} e^{i\hat{H}T} $
and then use the cyclic property of the trace.
We can also regard $e^{-\beta \hat{H}}$
as a time translation in the imaginary direction.
Then (\ref{ZJ2}) can be rewritten as
\bea
&&
\int d\Phi
\int d\Phi'
\int d\Phi''
\int d\Phi'''
\nn
&&\qquad\quad\,
\langle \Phi,-T| 
T_{C_1}[e^{i \int_{C_1} dt J(t)\hPhi(t)}]
|\Phi',T\rangle \label{C1}\\
&&\qquad\qquad
\langle \Phi',T| 
T_{C_3}[e^{i \int_{C_3} dt J(t)\hPhi(t)}]
|\Phi'',T-i\sigma \beta\rangle 
\label{C3}\\
&&\qquad\!\!\!
\langle \Phi'',T- i\sigma \beta|
T_{C_2}[e^{i \int_{C_2} dt J(t)\hPhi(t)}]
|\Phi''',-T-i\sigma \beta\rangle \label{C2}\\
&&\quad\!\!
\langle \Phi''',-T-i \sigma\beta| 
T_{C_4}[e^{i \int_{C_4} dt J(t)\hPhi(t)}]
|\Phi,-T-i\beta \rangle \label{C4}
\eea
where $|\hPhi,z \rangle$
is the eigen-state of the Heisenberg operators:
\bea
\hat{\Phi}(z)| \Phi,z \rangle
= 
{\Phi}| \Phi,z \rangle , \quad
| \Phi,z \rangle
= e^{i\hat{H} z} | \Phi\rangle,
\eea
and $z$ can be a complex number.
We have matched 
the eigen-states of the
Hisenberg operators with 
the eigen-states of the 
Shr\"odinger operators at $z=0$.
The projections to the physical states (\ref{phys})
is understood in the above expression.
When $\sigma = \frac{1}{2}$,
the time evolution
on
the vertical part of the contour
(\ref{C2})
is essentially equivalent
to preparing the thermal vacuum (\ref{vacb})
of the thermo-field dynamics.
(We will take $T\rightarrow \infty$ 
and
$J(t) \rightarrow 0$
as $t \rightarrow \pm \infty$.
To see the above, one just needs to change the notation:
One changes ``ket" of the
matrix elements of (\ref{C3}) to ``bra",
and regard them as states in doubled Hilbert space.
Then the matrix elements of (\ref{C1}) and (\ref{C2})
are combined into matrix elements acting on
this doubled Hilbert space.
Similar for the matrix elements of (\ref{C4}).)
By further inserting a complete set of physical states
satisfying (\ref{phys}),
we obtain the path integral representation
of the generating functional $Z[J]$:
\bea
 \label{PI}
Z[J]
= 
\int [{\cal D}_C \Phi] \,
e^{i \int_C dt
\left\{
{\cal L}[\Phi] + J(t)\Phi(t)
\right\}},
\eea
where ${\cal L}[\Phi]$ is the Lagrangian%
\bea
{\cal L}[\Phi] =
\frac{1}{g^2}
\tr
\left\{ 
\frac{1}{2}D_t \Phi D_t \Phi
-\frac{\w^2}{2} \Phi^2 - V[\Phi]
\right\} .
\eea
The covariant derivative is given by
\bea
(D_t \Phi )_{ab}
=
\pa_t \Phi_{ab}
- i [A_0,\Phi]_{ab},
\eea
and the gauge field $A_0$ has been introduced
while imposing the Gauss' law constraints
as delta function.
The path integral is over the fields
which satisfy the boundary conditions
\bea
 \label{bc}
\Phi_{ab}(-T-i\beta)
=
\Phi_{ab}(-T)  ,
\eea
following from the trace over
the Hilbert space in (\ref{TG}).
We can rewrite (\ref{PI}) as
\bea
 \label{ZJ}
Z[J]
= 
\exp
\left\{
{-i \int_C dt 
V\left[\frac{1}{i}\frac{\delta}{\delta J}\right]}
\right\}
\exp
\left\{
{-\frac{i}{2} g^2 \int_C dt \int_C dt'
J_{ab}(t) D^C_{ab,cd}(t-t') J_{cd}(t')}
\right\},\nn
\eea
where the thermal propagator
$D^C_{ab,cd}(t-t')$
is a Green's function on the contour
\bea
 \label{green}
(- \pa_t^2 - \w^2) D^C_{ab,cd} (t-t')
=
\delta_C (t-t') \delta_{ad} \delta_{bc} ,
\eea
subject to the boundary condition
following from (\ref{bc}).
Here $\delta_C (t-t')$ is the delta function
defined on the contour:
\bea
\int_C dt \delta_C (t-t') f (t) = f(t').
\eea
The boundary of the time $T$
is eventually taken to infinity.
By taking $J(t) \rightarrow 0$ as $t \rightarrow \pm \infty$,
the generating functional factorizes as
\bea
 \label{facto}
Z[J] = Z_{12}[J] Z_{34}[J] ,
\eea
where $Z_{12}[J]$ (respectively
$Z_{34}[J]$) denotes the contribution
from the path $C_1$ and $C_2$ ($C_3$ and $C_4$).
The effect of finite temperature enters
in the propagators through the
boundary condition (\ref{bc}).
Although the generating functional can be
seen to factorize, $Z_{34}[J]$ part plays a role
for modifying the boundary conditions on
the Green's function, as we will see below.

\subsection{Incorporating the effect of 
the confined phase background}

In this subsection we present a prescription
for reading off
the dual bulk description corresponding
to the confined phase
in the real time formalism.
When there are no external operator insertions,
one can take the Matsubara contour, i.e.
the line straight down from $-T$ to $-T-i\beta$.
Then the calculation reduces to that in the imaginary
time formalism.
As we have seen in section \ref{DecW},
the confined phase is characterized
by the vanishing
of the expectation value of the Polyakov loop.
The large $N$ saddle point value of the temporal
gauge field $A_0$ was given by
\bea
 \label{A0}
\frac{\pa}{\pa \tau} A_{0ab} = 0,\quad 
A_{0ab}
=
\delta_{ab}
\frac{2\pi}{\beta N}
\left( a-\frac{N+1}{2} \right),
\eea
in an appropriate gauge.
(\ref{A0}) corresponds to the
constant eigen-value density 
$\rho(\theta)=\frac{1}{2\pi}$ 
in the large $N$ limit
(see section \ref{DecW}).
This gives an appropriate expansion point
for perturbative calculation
on the vertical parts of the contour.

As we have seen in the previous section,
in the imaginary time formalism
it was essential to
expand around the saddle point of $A_0$
(\ref{A0})
to read off the dual bulk geometry in the confined phase.
Therefore, also in the real time formalism,
the correct prescription
for reading off the dual bulk description
corresponding to the confined phase
should be to include the saddle point value (\ref{A0})
of $A_0$
into the Green's function
{\em on the vertical parts of the contour}.
Thus, instead of (\ref{green}),
we use Green's function which satisfies
\bea
 \label{boxA}
(-D_t^2 - \w^2) D^C_{ab,cd} (t-t')
=
\delta_C (t-t') \delta_{ad} \delta_{bc} \quad ,
\eea
where on the vertical parts of the contour
we have included the saddle point value of
$A_0$ (\ref{A0})
in the covariant derivative $D_t$.
On the horizontal parts of the contour
one can choose $A_0 = 0$ gauge.
Since we are including
the effect of the $A_0$ configuration (\ref{A0})
on the vertical parts of the contour,
it is convenient to
define the field
\bea
 \label{redf}
\tilde{\Phi}_{ab}(t)
=
e^{-2\pi i\frac{a-b}{\beta N} (\mbox{\footnotesize Im}\, t)}
\Phi_{ab}(t)
\eea
so that the differential equation for the
Green's function $\tilde{D}^C(t-t')$ for $\tPhi(t)$
takes the form of the ordinary one (\ref{green}):
\bea
 \label{tbox}
(-\pa_t^2 - \w^2) \tilde{D}^C_{ab,cd} (t-t')
=
\delta_C (t-t') \delta_{ad} \delta_{bc}  \quad .
\eea
However, the field redefinition (\ref{redf})
modifies the boundary condition (\ref{bc}) to
\bea
 \label{tbd}
\tilde{\Phi}_{ab}(-T-i\beta)
=
e^{2\pi i\frac{a-b}{N}}
\tilde{\Phi}_{ab}(-T) .
\eea
One can solve (\ref{green}) with the ansatz
\bea
\tilde{D}^C_{ab,cd} (t-t')=
\theta_C(t-t') \tilde{D}^>_{ab,cd} (t-t')
+
\theta_C(t'-t) \tilde{D}^<_{ab,cd} (t-t') ,
\eea
where $\theta_C(t-t')$ is the step function
defined on the contour:
\bea
\theta_C(t-t')
= \int_C^t dt'' \delta_C(t''-t').
\eea
Since from
(\ref{PI}) to (\ref{ZJ})
the change of variable
\bea
\tPhi_{ab}(t) \rightarrow
\tPhi_{ab}(t) + \int_C dt' \tD_{ab,cd}(t-t') J_{cd}(t'),
\eea
has been made,
the boundary condition (\ref{tbd}) implies
\bea
 \label{bcD}
\tilde{D}^>_{ab,cd} (t-t'-i\beta)
=
e^{2\pi i\frac{a-b}{N}} \tilde{D}^<_{ab,cd} (t-t') .
\eea
The unique solution to (\ref{tbox})
with the boundary condition (\ref{bcD}) is
\bea
 \label{DABCD}
&&\tilde{D}^C(t-t')_{ab,cd}
=
\frac{-i}{2\w}
\left[
(A e^{-i\w t} + B e^{i\w t}) \theta_C(t-t')
+
(C e^{-i\w t} + D e^{i\w t}) \theta_C(t'-t)
\right] \nn
\eea
with
\bea
 \label{ABCD}
A =
\frac{1}{1 - e^{-\beta \w - 2\pi i \frac{a-b}{N}}} ,
&&
B =
\frac{e^{-\beta \w + 2\pi i \frac{a-b}{N}}}%
{1 -  e^{-\beta \w + 2\pi i \frac{a-b}{N}}} , \nn
C =
\frac{e^{-\beta \w - 2\pi i \frac{a-b}{N}}}%
{1 -  e^{-\beta \w - 2\pi i \frac{a-b}{N}}}  ,
&&
D =
\frac{1}{1 - e^{-\beta \w + 2\pi i \frac{a-b}{N}}} .
\eea
The Green's function
(\ref{DABCD}) can be rewritten in the
spectral representation:
\bea
 \label{tDC}
i \tilde{D}^{C}_{ab,cd} (t-t')
=
\int_{-\infty}^\infty \frac{dk_0}{2\pi}
e^{-i k_0 (t-t')}
\rho (k_0)
[\theta_C(t-t')+N(k_0,a-b)]\delta_{ad} \delta_{bc}
\quad ,
\eea
where
\bea
\rho (k_0) = 2\pi \ve(k_0) \delta(k_0^2 -\w^2)
, \quad \ve(k_0) = \theta(k_0) - \theta(- k_0)
\eea
and
\bea
\label{Nk}
N(k_0,a-b)
= \frac{1}{e^{\beta k_0 + 2\pi i \frac{a-b}{N}} - 1}.
\eea
As in (\ref{facto}),
the partition function 
factorizes.
Therefore, only
the propagators between the fields
on the contours $C_1$ or $C_2$
need to be considered.
The propagators for general $\sigma$
($0< \sigma < 1$, where
$\sigma$ is given in Fig.\ref{contour})
are obtained as
\bea
\tD^{(11)}_{ab,cd}(t-t')
&=& \tD^C_{ab,cd}(t-t'), \label{D11}\\
\tD^{(22)}_{ab,cd}(t-t')
&=& \tD^C_{ab,cd}((t-i\sigma\beta)-(t'-i\sigma\beta)), \label{D22}\\
\tD^{(12)}_{ab,cd}(t-t')
&=& \tD^<_{ab,cd}(t-(t'-i\sigma\beta)), \label{D12}\\
\tD^{(21)}_{ab,cd}(t-t')
&=& \tD^>_{ab,cd}((t-i\sigma\beta)-t')\label{D21}.
\eea
Notice that the propagator takes the form of
a $2\times 2$ matrix.
This can be looked as the degrees of freedom
are doubled compared with the original theory
at zero temperature.
The doubling of the degrees of freedom
originates from the two parts of the
contour $C_1$ and $C_2$
in Fig.\ref{contour}.
$\tD^{(11)}$ (respectively $\tD^{(22)}$) is 
regarded as a propagator between
type-1 (type-2) fields, and
$\tD^{(12)}$ and $\tD^{(21)}$ are mixed
propagators
between type-1 and type-2 fields.


By taking $\sigma = \frac{1}{2}$
we obtain the most symmetric expression.
It is convenient to split the
propagator into a temperature dependent part
and an independent part.
Also, at this point it is convenient
to undo the field redefinition
(\ref{redf})
to obtain a symmetric expression
for the propagators.
In momentum space,
they are given by
\bea
 \label{dvD}
i D^{(rs)}_{ab,cd}
=
i D^{(rs)}_{0 ab,cd}  +
i D^{(rs)}_{\beta ab,cd}
\quad (r,s = 1,2),
\eea
\bea
 \label{D0}
i D_{0 ab,cd} =
\delta_{ad}\delta_{bc}
\left(
\begin{array}{cc}
 \frac{i}{k_0^2-\w^2+i\e} & 0 \\
 0 & \frac{-i}{k_0^2-\w^2-i\e}
\end{array}
\right) ,
\eea
\bea
 \label{Db}
i D_{\beta ab,cd} &=&
\delta_{ad}\delta_{bc}
\pi \delta(k_{0}^2- \w^2) \nn
&&
\times
\frac{1}{e^{{|\beta k_0+2\pi i\frac{a-b}{N}}|_R}-1}
\left(
\begin{array}{cc}
 1 & e^{\frac{1}{2}|\beta k_{0}+2\pi i\frac{a-b}{N}|_R} \\
e^{\frac{1}{2}|\beta k_{0}+2\pi i\frac{a-b}{N}|_R} & 1
\end{array}
\right) .
\eea
In the above, $|\cdots |_R$ is defined as
\bea
 \label{||R}
|z|_R =
\left\{
\begin{array}{c}
z \quad (\mbox{Re}\, z > 0)\\
-z \quad (\mbox{Re}\, z < 0)
\end{array}
\right.  , \quad (\mbox{Re}\, z \ne 0).
\eea
Eq.(\ref{||R}) is not defined for
$\mbox{Re}\, z = 0$, but
because
of the on-shell delta function in
(\ref{Db}) one does not need to consider
that case as long as $\w \ne 0$.
Notice the gauge index dependant
phases in (\ref{Db}) which arouse
from our prescription for
incorporating the effect of the confined phase
background. 
These will play the crucial roles
in the following discussions.
\begin{figure}
\begin{center}
 \leavevmode
 \epsfxsize=95mm
 \epsfbox{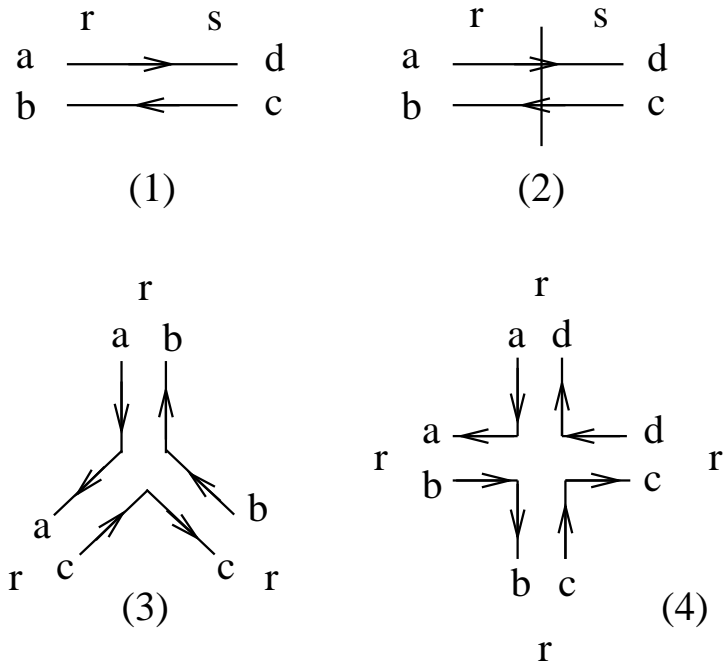}\\
\end{center}
\caption{Feynman rules for the real time formulation of
large $N$ gauge theories at finite temperature:
(1) is the temperature independent part
of the propagator (\ref{D0}), while
(2) is the temperature dependent part
(\ref{Db}).
The temperature dependent part
of the propagator (2) is drawn with
a ``cut" line (the vertical line in the figure).
The interaction vertices (3),(4) are drawn
schematically, just to show the index flow structure.
The interaction vertices do not mix the type-1 fields
with the type-2 fields.}
\label{Feyn}
\end{figure}

Perturbative Feynman rules can be obtained
just as in the conventional field theories
and are sketched in Fig.\ref{Feyn}.
(1) represents the temperature independent
part of the propagator $iD^{(rs)}_{0 ab,cd}$
and (2)
the temperature
dependent part
$iD^{(rs)}_{\beta ab,cd}$.
The temperature dependent part
of the propagator is drawn with
the ``cut" line in Fig.\ref{Feyn} (2).
{This cut is one of the most important
tools we will use repeatedly
in the following discussions.}
Type-1 fields and type-2 fields
are coupled only through the propagators:
The interaction vertices do not mix
type-1 and type-2 fields.
The interaction vertices of type-2 fields
($\times i$) are given by the complex conjugate
of those of type-1 fields:
\bea
 \label{rp}
i\, \tr V_2[\Phi_{(2)}]
=
\left.(i\, \tr V_1[\Phi_{(1)}])^*
\right|_{\Phi_{(1)}\rightarrow \Phi_{(2)}}.
\eea
We have assumed that the potential is real.

\subsection{The vanishing mechanism and surviving diagrams
as real time closed string diagrams}

In this subsection,
with the prescription for incorporating
the effect of
the confined phase background (\ref{A0})
discussed in the previous subsection,
we will show that
the contributions from a large class of
Feynman diagrams vanish.
%
The quantities of our interest
are the correlation functions
of gauge invariant
single trace local operators,
which correspond to closed string states
in the AdS-CFT correspondence.
Throughout this section we will work
in the planar limit $g \rightarrow 0$,
$N \rightarrow \infty$ with
the 't Hooft coupling $g^2 N$ fixed.
In the planar limit,
one can always associate
a loop momentum to an index loop,
as will be explained below.%
\footnote{There are $\ell + 1$ index loops for
$\ell$ (momentum) loop planar diagrams
of a correlation function of
gauge invariant operators,
but one summation over gauge indices
decouples since the gauge indices always
appear as a difference of two indices
\cite{Furuuchi:2005qm,Furuuchi:2005eu}.}
The total momentum on a propagator
is a sum of two momenta
associated with the index lines
(taking into account the sign indicated by the arrows),
and an external momentum flow
if there is any.
By a shift of loop momenta, which are
integration variables,
one can choose any tree sub-diagram
connecting the external legs
to express the external momentum flow.
But once it is chosen,
the integrations over loop momenta
should be done
with
that fixed external momentum flow.

We first consider Feynman diagrams which have
at least one index loop containing
only one cut (which is denoted as $a_i$ below).
\begin{figure}
\begin{center}
 \leavevmode
 \epsfxsize=75mm
 \epsfbox{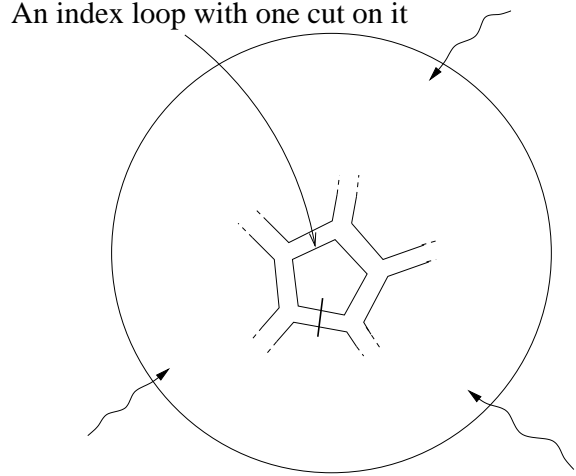}\\
\end{center}
\caption{A 't Hooft-Feynman diagram 
with an index loop with only one cut on it.}
\label{1cut}
\end{figure}
In this case the cut must be either
1-1 or 2-2 cut.\footnote{When there is a 1-2 cut
on an index loop, there must be a 2-1 cut
on that index loop.} 
By a shift of loop momenta,
without loss of generality 
one can draw 
the flow of the external momentum
without crossing the cut.
From (\ref{Db}),
the diagram with the 1-1 or 2-2 cut
is proportional to a factor
\bea
 \label{phase}
\sum_{a_i=1}^N
\frac{1}{e^{|\beta(p_{0i}-p_{0j})
+2\pi i \frac{a_i-a_j}{N}|_R}-1}.
\eea
Here, $i,j$ label the index-momentum loops.
As mentioned earlier,
the loop momentum $p_{0i}$
is ``associated" with the gauge index $a_i$,
that means, they always appear
in the combination
$\beta p_{0i}
+2\pi i \frac{a_i}{N}$.
The origin of this combination
is the covariant derivative for
adjoint fields.
Therefore, by taking the background gauge
around the field configuration (\ref{A0}),
even when there are derivative couplings
the loop momentum and the associated index
also appear in the same combination.
Such derivative couplings give rise
to a multiplicative factor
which is polynomial in
$\beta p_{0i}+2\pi i \frac{a_i}{N}$.
Those just require a minor modification
in the following discussions
and do not change the conclusion about
whether a diagram vanishes or not.
Therefore, to keep the essential points clear
in the presentation,
we will only write down the formula
for the case in which such
derivative couplings are absent.
Since we are working in the strict $N \rightarrow \infty$
limit,
the sum over
the gauge indices
$a_i$ can be replaced by the integral:
$\frac{a_i}{N} \rightarrow \theta_i$,
$\sum_{a_i=1}^N \rightarrow N \int_0^1 d\theta_i$.
(To avoid repetition,
this replacement will be
implicit in what follows.)
Then, one can Fourier expand
the integrand of (\ref{phase}) as
\bea
\label{vmec}
\int_0^1 d\theta_i
\frac{1}{e^{|\beta(p_{0i}-p_{0j})
+2\pi i (\theta_i-\theta_j)|_R}-1}
&=&
\int_0^1 d\theta_i
\sum_{n=1}^\infty
e^{-n |\beta (p_{0i}-p_{0j})
+2\pi i (\theta_i-\theta_j)|_R} \nn
&=& 0 .
\eea
From the definition (\ref{||R}),
this kind of diagram
has either {all} negative (when $p_{0i}>p_{0j}$)
or {all} positive (when $p_{0i}<p_{0j}$)
powers of $e^{2\pi i \theta_i}$.
In either case,
(\ref{vmec}) vanishes.
{\em Eq.(\ref{vmec}) is the basic equation
relevant for
selecting the non-vanishing Feynman diagrams
in the confined phase}.

\begin{figure}
\begin{center}
 \leavevmode
 \epsfxsize=100mm
 \epsfbox{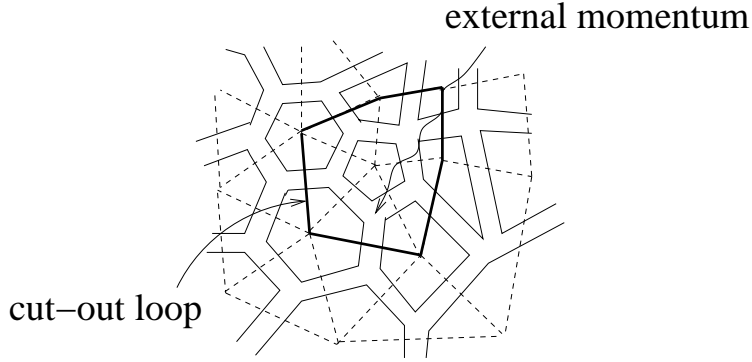}\\
\end{center}
\caption{A cut-out loop on the dual graph of
a 't Hooft-Feynman diagram.}
\label{cutdual}
\end{figure}

\begin{figure}
\begin{center}
 \leavevmode
 \epsfxsize=70mm
 \epsfbox{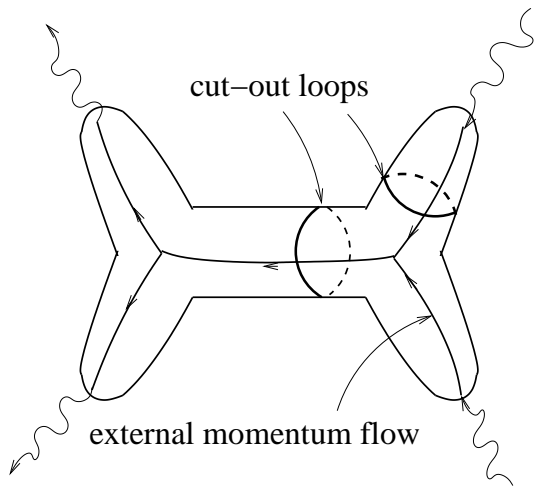}\\
\end{center}
\caption{A `t Hooft-Feynman diagram
identified with a closed string worldsheet.
The cut-out loop must cross the tree sub-diagram
where the external momentum flows.
Thus the cut-out loop can be identified
with the cut on a closed string propagator.}
\label{tree}
\end{figure}

\begin{figure}
\begin{center}
 \leavevmode
 \epsfxsize=70mm
 \epsfbox{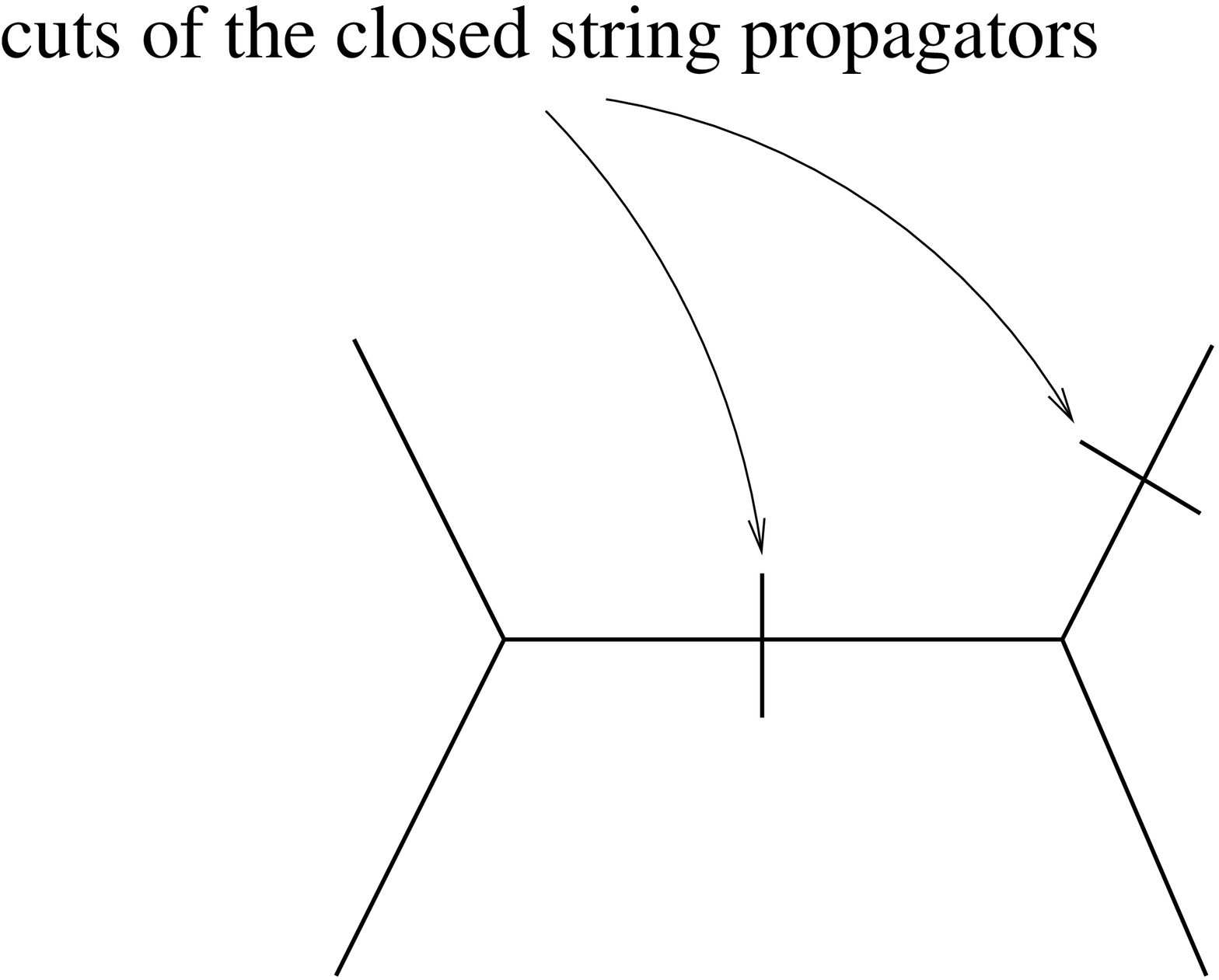}\\
\end{center}
\caption{Fig.\ref{tree} re-drawn
as a closed string tree diagram 
in the real time formalism. 
The cut-out loops in Fig.\ref{tree} are
identified with the cuts on the closed string propagators.}
\label{ctree}
\end{figure}

The above vanishing mechanism
means that
if there is a cut ``coming
into" a face, there must be another
cut ``going out".
More precisely,
it gives a conservation law for 
the phases 
($2\pi \theta_i$ in (\ref{vmec}))
at a face of the diagram,
or a vertex in the dual graph.
Thus in general, on the
surviving diagrams
those cuts must
make up closed circuits
on the dual graph (Fig.\ref{cutdual}).
We have connected the starting and end points
of the cuts on the same face of the diagram,
because those faces are
to be ``filled" to make up
closed string worldsheet.
Those closed circuits
divide the worldsheet
into regions which do not contain
further cuts, otherwise 
the diagram will vanish (Fig.\ref{tree}).
One can regard
those closed circuits
as made of
``cut-out loops" \cite{Furuuchi:2005zp}.
For a closed string interpretation to be valid,
the temperature
dependence should appear in this way,
since if the temperature dependence 
appears without making a closed loop
on the diagram,
it means that we are not just seeing 
the closed string
but rather its sub-structure.
(However, this is a description 
based on the expansion around
the AdS geometry in the bulk.
As we have seen in the previous section
in the Euclidean case,
in order for the closed string interpretation
to be valid, the temperature dependence should
be absorbed into the change of geometry 
in the bulk space-time.)
One can show that
the external momentum flow
must cross the cut-out loop \cite{Furuuchi:2005zp}.
Since the 
external momentum flow can be drawn on a
tree sub-diagram of the 't Hooft-Feynman diagram,
each cut-out loop
can be associated with
an edge of the closed string tree diagram
(Fig.\ref{ctree}).
This indicates that
the 't Hooft-Feynman diagrams of the
Yang-Mills theory at finite temperature
in the confined phase
give rise to 
closed string Feynman diagrams in a
closed string field theory
in the real time formalism.

One can also show
that the cut has
a correct energy dependence
to be identified with 
a cut in the closed string field theory propagator
in the real time formalism \cite{Furuuchi:2005zp}.\footnote{%
It was shown in the free field limit but
expected to hold for finite coupling.} 
This is what the cuts 
in the picture should actually mean.
The energy dependence
of the cut in general has a form
of eq.(\ref{Db}) without the gauge index dependent
phase factor which is specific to our model,
as one can see from the generality
of the derivation.

\begin{figure}
\begin{center}
 \leavevmode
 \epsfxsize=110mm
 \epsfbox{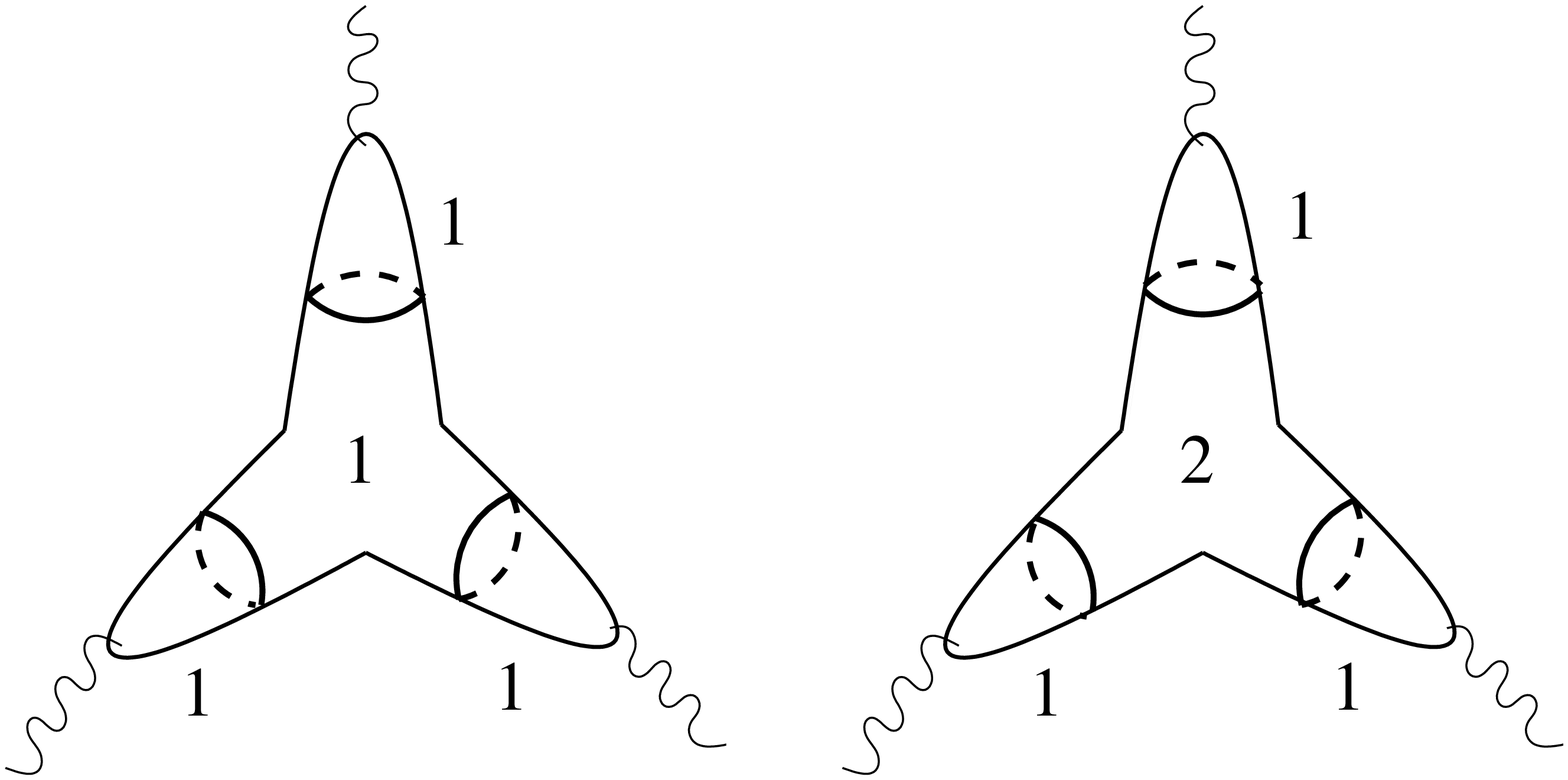}\\
\end{center}
\caption{'t Hooft-Feynman diagrams divided
into type-1 regions and type-2 regions
by the cut-out loops.
The type-2 region in the diagram on the right
is complex conjugate to the corresponding
type-1 region in the diagram on the left.}
\label{closedv}
\end{figure}

Furthermore,
we see that
the propagator
and interaction vertices ($\times i$)
of the
type-2 string fields 
of this hypothetical closed string field theory
are complex conjugate
of those of type-1.
This is the property
that in general
field theory in the real time formalism
should satisfy,
see eqs.(\ref{D0}),
(\ref{Db}),\footnote{Again, the gauge index dependent
phases in eq.(\ref{Db}) are specific to
our gauge theory model and usually do not appear, 
and cancel in the surviving
diagrams in the confined phase as we have seen.}
and (\ref{rp}).
The surviving
't Hooft-Feynman diagrams
are divided into regions
inside which there's no more cut
(otherwise it vanishes as we have seen).
Since the cut is the only temperature
dependent part, this means that
each region probes the same geometry
as the original zero-temperature theory.
Since type-1 fields and type-2 fields
mixes only through the cuts,
each region contains either 
only type-1 propagators and interaction vertices
or those of only type-2.
For every diagram
with a region made of type-2,
there must be a
corresponding diagram with
all propagators and vertices in that
region are replaced by those of type-1
(Fig.\ref{closedv}).
This leads to the conclusion that
the type-2 propagators
and vertices of 
closed string field theory
is complex conjugate of
those of type-1.

Thus,
the dual description
of the Yang-Mills theory
in the confined phase
can be consistently identified
with 
{\em the
closed string field theory
at finite temperature
in the real time formalism
whose target space is the same as
that of the zero-temperature theory}.%
\footnote{The main reason we used 
string ``field" theory here
is that 
the worldline/worldsheet
formulation
of finite temperature field
theory in the real time formalism
is not developed enough.
However, see \cite{Mathur:1993tp} for some
investigation of the first-quantized approach.
We do not intend to 
persue the non-perturbative analysis,
which was one of the motivations in 
the construction of string field theory.
We are just comparing
the perturbative diagrams we would obtain
from the field theory.}
The field contents and 
the target space geometry
in the bulk,
obtained from the Yang-Mills side,
are
all uplifted from
those obtained from the Yang-Mills theory
at zero-temperature,
just like 
in the standard real time formulation of
field theory
at finite temperature.
This can be regarded as a Lorentzian
version of the large $N$ reduction.
The point is that one can use
most of the results in the zero-temperature
also at finite temperature
in the confined phase.
If our understanding of the
AdS-CFT correspondence 
at zero-temperature becomes more precise,
so as the case at finite temperature
in the confined phase.

\subsection{Comments on the deconfined phase
and the black hole geometry}

\begin{figure}
\begin{center}
 \leavevmode
 \epsfxsize=35mm
 \epsfbox{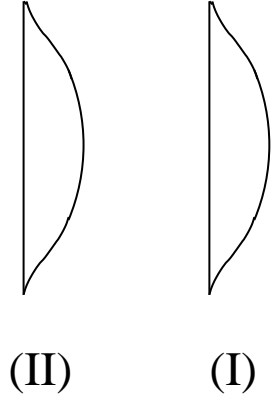}\\
\end{center}
\caption{The Carter-Penrose diagram
of the AdS space (I) and its copy (II).}
\label{2AdS}
\end{figure}

\begin{figure}
\begin{center}
 \leavevmode
 \epsfxsize=70mm
 \epsfbox{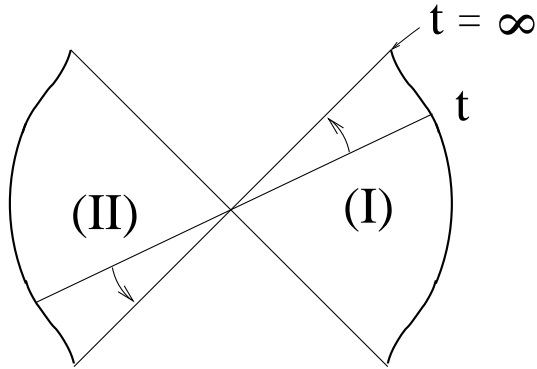}\\
\end{center}
\caption{The Carter-Penrose diagram
of the AdS-Schwarzschild black hole geometry
with the asymptotic static observer time.
The region inside the black hole (compare
with Fig.\ref{AdSBH}) is not reachable
in finite asymptotic observer time $t$.}
\label{BHasy}
\end{figure}

One might have 
had an impression that
the structure of the bulk space-time
are
very different in the confined phase 
and the deconfined phase,
because the one has a black hole
and the other does not.
This is correct, but
we should also point out the similarity
in the two descriptions.
As one can see from the derivation
of the path integral formulation of the
real time formalism,
one can either regard
the real time formulation
in AdS
as two kinds of fields living in the same AdS,
or instead
the same fields living in two copies of 
AdS ((I) and (II) in Fig.\ref{2AdS}).
In a similar way,
in the black hole phase
one can regard the
fields to live on copies
region I and region II
in Fig.\ref{BHasy}.
Although in the 
maximally
extended AdS-Schwarzschild geometry
these regions are connected
through the region
inside the black hole
(the upper and lower triangle regions containing
the singularity in Fig.\ref{AdSBH}.\footnote{%
We have called the region (II) 
``behind" the horizon since when looked
from the region (I) it is so,
but it is outside the another horizon
and hence it is ``outside" the black hole.}),
the Yang-Mills theory does not
see these region,
at least in a naive way,\footnote{%
Nevertheless researchers
challenge for extracting
the information inside the black hole.
See e.g.
\cite{Kraus:2002iv,Fidkowski:2003nf,Festuccia:2005pi} 
for such efforts.
This may be possible if there is
a black hole complementarity \cite{Susskind:1993if},
but at the same time
it might make the
description inside the black hole
redundant,
especially if one regards
CFT side more fundamental and
the space-time concept in the bulk
as emerging.}
because
the boundary time corresponds to
the asymptotic static observer's time
which does not cross 
the horizon (see Fig.\ref{BHasy}).
If one looks at only the region (I) and (II)
in Fig.\ref{BHasy},
one may recognize that for the asymptotic observer
it is the same
real time formalism
applied to the different geometries, i.e.
the AdS geometry and the black hole geometry.
Recall that 
the vertical parts of the contour
which we sent to $t=\pm \infty$ 
put boundary conditions for the propagator
when we derived
our perturbative Feynman rules.\footnote{%
As we have mentioned in subsection \ref{PIr},
the time translation in the vertical part
of the contour can be identified with
the thermal vacuum in thermo-field dynamics.
The thermal vacuum
(\ref{vacb}) in the deconfined phase
should correspond to choosing the
Hartle-Hawking vacuum
which is obtained
by time translation
in the Euclidean section 
of the black hole geometry and is invariant
under the isometry generated by $\pa_t$
in the bulk
\cite{Maldacena:2001kr}.
(For a description of the Hartle-Hawking vacuum
related the discussion here, see \cite{Jacobson:1994fp}.
See e.g. \cite{Wald:1995yp} for 
a examination of the Hartle-Hawking vacuum
in a physically realistic context.) 
This should be understood if one recalls
that in the Hamiltonian formulation,
AdS-CFT correspondence is a (conjectural)
isomorphism between Hilbert spaces
of closed string theory on AdS
and boundary CFT.
The thermo-field dynamics on the boundary
should essentially be the dual of the
Israel's description of
Hawking radiation
via the thermo-field dynamics
\cite{Israel:1976ur}.}


In the deconfined phase,
the non-singlet states 
which cannot be
interpreted as closed strings
become important.\footnote{See
\cite{Maldacena:2005hi} for a study
of a non-singlet sector 
in a two-dimensional model in the Lorentzian signature.}
It is tempting to identify
those non-singlet degrees of freedom
with closed strings which look like 
open strings half hidden in the black hole horizon,
but this interpretation requires 
further clarification.

It may be good to keep in mind that
the appearance of the non-singlet states
indicates that
the bulk space-time picture
probed by closed strings might be missing
some important 
information of the bulk,
especially 
about the black hole.

\section*{Acknowledgments}
This note is 
based on talks presented at
IPM String School and Workshop April 2006,
Strings 2006 Shanghai Workshop 
and Hangzhou Workshop June
as well as lectures derivered at
ICTS-USTC March 2006
and
at Chulalongkorn University April 2006.
I would like to thank the organizers 
for the very nice workshops
and the participants/attendants
for stimulating discussions.
I am also benefitted from the discussions
at Strings 2006 Beijing and would like to
thank for the support.
I am also benefitted from
my stay in NTNU and KIAS
as well as visits to
NCTS, NTU, CQUeST, Tokyo U.(Komaba), 
KEK, Osaka U. and Kyoto U..
I would like to thank
D. Astefanesei,
J. R. David,
D. Ghoshal,
K. Hashimoto,
C. A. R. Herdeiro,
T. Hirayama, 
P.-M. Ho,
Y. Hosotani,
N. Iizuka,
D. P. Jatkar,
S. Kalyana Rama,
Y. Kazama,
T. Kimura,
A. Lawrence,
K.-M. Lee,
F.-L. Lin, 
H. Liu,
J.-X. Lu,
T. Matsuo,
S. Naik,
S. Nakamura,
T. Nakatsu,
J. Nishimura,
B. Sathiapalan,
E. Schreiber,
G. W. Semenoff,
A. Sen,
M. Shigemori,
S.-J. Sin,
T. Takayanagi,
S. Teraguchi,
S. Terashima,
D. Tomino,
S. Wadia, 
P. Yi, 
K. P. Yogendran,
T. Yoneya,
A. Zamolodchikov 
and especially R. Gopakumar
for useful discussions, explanations, 
comments and encouragements.
I would also like to thank
Ms. Ma,
B. Ning,
Z.-L. Wang, 
R.-J. Wu, 
M. Alishahiha, 
R. Fareghbal,
M. M. Sheikh-Jabbari,
Ms. Shirine,
N. Salehirad,
A. Chatrabhuti,
R. Dhanawittayapol,
A. Ungkitchanukit,
P. Ittisamai,
P. Karndumri,
P. Patcharamaneepakorn,
P. Wongjun,
P. Kirdruang,
S. Sunanta,
C.-C. Hsieh,
F.-L. Lin,
L.-L. Tsai,
P. Chingangbam,
S. Dutta, 
B.-H. Lee,
H.j. Shin,
S.-H. Yoon,
S. Hu,
S. G. Nibbelink,
Ms. Shen,
X. Xie,
B. Zhao,
S. S. Deshingkar,
N. Mahajan,
A. K. Ray,
and 
many students, 
postdocs, 
professors,
secretaries,
local assistants and friends
whose names I could not list here
for warm hospitality and kind help.
Last but not least,
I would like to thank all the friends 
in my home institute
for their continuous support.

\bibliography{lec}

\providecommand{\href}[2]{#2}\begingroup\raggedright\begin{thebibliography}{10}

\bibitem{Maldacena:1997re}
J.~M. Maldacena, ``The large N limit of superconformal field theories and
  supergravity,'' { Adv. Theor. Math. Phys.} {\bf 2} (1998) 231--252,
\href{http://www.arXiv.org/abs/hep-th/9711200}{{hep-th/9711200}}.

\bibitem{Aharony:1999ti}
O.~Aharony, S.~S. Gubser, J.~M. Maldacena, H.~Ooguri, and Y.~Oz, ``Large N
  field theories, string theory and gravity,'' { Phys. Rept.} {\bf 323} (2000)
  183--386,
\href{http://www.arXiv.org/abs/hep-th/9905111}{{hep-th/9905111}}.

\bibitem{Sundborg:1999ue}
B.~Sundborg, ``The Hagedorn transition, deconfinement and N = 4 SYM theory,'' {
  Nucl. Phys.} {\bf B573} (2000) 349--363,
\href{http://www.arXiv.org/abs/hep-th/9908001}{{hep-th/9908001}}.

\bibitem{Polyakov:2001af}
A.~M. Polyakov, ``Gauge fields and space-time,'' { Int. J. Mod. Phys.} {\bf
  A17S1} (2002) 119--136,
\href{http://www.arXiv.org/abs/hep-th/0110196}{{hep-th/0110196}}.

\bibitem{Aharony:2003sx}
O.~Aharony, J.~Marsano, S.~Minwalla, K.~Papadodimas, and M.~Van~Raamsdonk,
  ``The Hagedorn / deconfinement phase transition in weakly coupled large N
  gauge theories,'' { Adv. Theor. Math. Phys.} {\bf 8} (2004) 603--696,
\href{http://www.arXiv.org/abs/hep-th/0310285}{{hep-th/0310285}}.

\bibitem{Aharony:2005bq}
O.~Aharony, J.~Marsano, S.~Minwalla, K.~Papadodimas, and M.~Van~Raamsdonk, ``A
  first order deconfinement transition in large N Yang- Mills theory on a small
  S**3,'' { Phys. Rev.} {\bf D71} (2005) 125018,
\href{http://www.arXiv.org/abs/hep-th/0502149}{{hep-th/0502149}}.

\bibitem{Gopakumar:2003ns}
R.~Gopakumar, ``From free fields to AdS,'' { Phys. Rev.} {\bf D70} (2004)
  025009,
\href{http://www.arXiv.org/abs/hep-th/0308184}{{hep-th/0308184}}.

\bibitem{Gopakumar:2004qb}
R.~Gopakumar, ``From free fields to AdS. II,'' { Phys. Rev.} {\bf D70} (2004)
  025010,
\href{http://www.arXiv.org/abs/hep-th/0402063}{{hep-th/0402063}}.

\bibitem{Gopakumar:2005fx}
R.~Gopakumar, ``From free fields to AdS. III,'' { Phys. Rev.} {\bf D72} (2005)
  066008,
\href{http://www.arXiv.org/abs/hep-th/0504229}{{hep-th/0504229}}.

\bibitem{Gopakumar:2004ys}
R.~Gopakumar, ``Free field theory as a string theory?,'' { Comptes Rendus
  Physique} {\bf 5} (2004) 1111--1119,
\href{http://www.arXiv.org/abs/hep-th/0409233}{{hep-th/0409233}}.

\bibitem{Aharony:2006th}
O.~Aharony, Z.~Komargodski, and S.~S. Razamat, ``On the worldsheet theories of
  strings dual to free large N gauge theories,'' { JHEP} {\bf 05} (2006) 016,
\href{http://www.arXiv.org/abs/hep-th/0602226}{{hep-th/0602226}}.

\bibitem{David:2006qc}
J.~R. David and R.~Gopakumar, ``From spacetime to worldsheet: Four point
  correlators,''
\href{http://www.arXiv.org/abs/hep-th/0606078}{{hep-th/0606078}}.

\bibitem{Yaakov:2006ce}
I.~Yaakov, ``Open and closed string worldsheets from free large N gauge
  theories with adjoint and fundamental matter,''
\href{http://www.arXiv.org/abs/hep-th/0607244}{{hep-th/0607244}}.

\bibitem{Furuuchi:2005qm}
K.~Furuuchi, ``From free fields to AdS: Thermal case,'' { Phys. Rev.} {\bf D72}
  (2005) 066009,
\href{http://www.arXiv.org/abs/hep-th/0505148}{{hep-th/0505148}}.

\bibitem{Furuuchi:2005zp}
K.~Furuuchi, ``Confined phase in the real time formalism and the fate of the
  world behind the horizon,'' { Phys. Rev.} {\bf D73} (2006) 046004,
\href{http://www.arXiv.org/abs/hep-th/0510056}{{hep-th/0510056}}.

\bibitem{Hawking:1982dh}
S.~W. Hawking and D.~N. Page, ``THERMODYNAMICS OF BLACK HOLES IN ANTI-DE SITTER
  SPACE,'' { Commun. Math. Phys.} {\bf 87} (1983)
577.

\bibitem{Witten:1998qj}
E.~Witten, ``Anti-de Sitter space and holography,'' { Adv. Theor. Math. Phys.}
  {\bf 2} (1998) 253--291,
\href{http://www.arXiv.org/abs/hep-th/9802150}{{hep-th/9802150}}.

\bibitem{Witten:1998zw}
E.~Witten, ``Anti-de Sitter space, thermal phase transition, and confinement in
  gauge theories,'' { Adv. Theor. Math. Phys.} {\bf 2} (1998) 505--532,
\href{http://www.arXiv.org/abs/hep-th/9803131}{{hep-th/9803131}}.

\bibitem{Gibbons:1976ue}
G.~W. Gibbons and S.~W. Hawking, ``ACTION INTEGRALS AND PARTITION FUNCTIONS IN
  QUANTUM GRAVITY,'' { Phys. Rev.} {\bf D15} (1977)
2752--2756.

\bibitem{Rey:1998ik}
S.-J. Rey and J.-T. Yee, ``Macroscopic strings as heavy quarks in large N gauge
  theory and anti-de Sitter supergravity,'' { Eur. Phys. J.} {\bf C22} (2001)
  379--394,
\href{http://www.arXiv.org/abs/hep-th/9803001}{{hep-th/9803001}}.

\bibitem{Maldacena:1998im}
J.~M. Maldacena, ``Wilson loops in large N field theories,'' { Phys. Rev.
  Lett.} {\bf 80} (1998) 4859--4862,
\href{http://www.arXiv.org/abs/hep-th/9803002}{{hep-th/9803002}}.

\bibitem{Rey:1998bq}
S.-J. Rey, S.~Theisen, and J.-T. Yee, ``Wilson-Polyakov loop at finite
  temperature in large N gauge theory and anti-de Sitter supergravity,'' {
  Nucl. Phys.} {\bf B527} (1998) 171--186,
\href{http://www.arXiv.org/abs/hep-th/9803135}{{hep-th/9803135}}.

\bibitem{Brandhuber:1998bs}
A.~Brandhuber, N.~Itzhaki, J.~Sonnenschein, and S.~Yankielowicz, ``Wilson loops
  in the large N limit at finite temperature,'' { Phys. Lett.} {\bf B434}
  (1998) 36--40,
\href{http://www.arXiv.org/abs/hep-th/9803137}{{hep-th/9803137}}.

\bibitem{Semenoff:2002kk}
G.~W. Semenoff and K.~Zarembo, ``Wilson loops in SYM theory: From weak to
  strong coupling,'' { Nucl. Phys. Proc. Suppl.} {\bf 108} (2002) 106--112,
\href{http://www.arXiv.org/abs/hep-th/0202156}{{hep-th/0202156}}.

\bibitem{Rohm:1983aq}
R.~Rohm, ``SPONTANEOUS SUPERSYMMETRY BREAKING IN SUPERSYMMETRIC STRING
  THEORIES,'' { Nucl. Phys.} {\bf B237} (1984)
553.

\bibitem{Sen:1998sm}
A.~Sen, ``Tachyon condensation on the brane antibrane system,'' { JHEP} {\bf
  08} (1998) 012,
\href{http://www.arXiv.org/abs/hep-th/9805170}{{hep-th/9805170}}.

\bibitem{KalyanaRama:1998cb}
S.~Kalyana~Rama and B.~Sathiapalan, ``The Hagedorn transition, deconfinement
  and the AdS/CFT correspondence,'' { Mod. Phys. Lett.} {\bf A13} (1998)
  3137--3144,
\href{http://www.arXiv.org/abs/hep-th/9810069}{{hep-th/9810069}}.

\bibitem{Barbon:2001di}
J.~L.~F. Barbon and E.~Rabinovici, ``Closed-string tachyons and the Hagedorn
  transition in AdS space,'' { JHEP} {\bf 03} (2002) 057,
\href{http://www.arXiv.org/abs/hep-th/0112173}{{hep-th/0112173}}.

\bibitem{Barbon:2002nw}
J.~L.~F. Barbon and E.~Rabinovici, ``Remarks on black hole instabilities and
  closed string tachyons,'' { Found. Phys.} {\bf 33} (2003) 145--165,
\href{http://www.arXiv.org/abs/hep-th/0211212}{{hep-th/0211212}}.

\bibitem{Barbon:2004dd}
J.~L.~F. Barbon and E.~Rabinovici, ``Touring the Hagedorn ridge,''
\href{http://www.arXiv.org/abs/hep-th/0407236}{{hep-th/0407236}}.

\bibitem{McGreevy:2005ci}
J.~McGreevy and E.~Silverstein, ``The tachyon at the end of the universe,'' {
  JHEP} {\bf 08} (2005) 090,
\href{http://www.arXiv.org/abs/hep-th/0506130}{{hep-th/0506130}}.

\bibitem{Horowitz:2006mr}
G.~T. Horowitz and E.~Silverstein, ``The inside story: Quasilocal tachyons and
  black holes,'' { Phys. Rev.} {\bf D73} (2006) 064016,
\href{http://www.arXiv.org/abs/hep-th/0601032}{{hep-th/0601032}}.

\bibitem{Furuuchi:2006st}
K.~Furuuchi, ``Matrix model for Polyakov loops, string field theory in the
  temporal gauge, and winding string condensation in anti-de Sitter space,''
\href{http://www.arXiv.org/abs/hep-th/0608108}{{hep-th/0608108}}.

\bibitem{Harmark:2006ta}
T.~Harmark and M.~Orselli, ``{Matching the Hagedorn temperature in AdS/CFT},''
  { Phys. Rev.} {\bf D74} (2006) 126009,
\href{http://www.arXiv.org/abs/hep-th/0608115}{{hep-th/0608115}}.

\bibitem{Aharony:1998qu}
O.~Aharony and E.~Witten, ``Anti-de Sitter space and the center of the gauge
  group,'' { JHEP} {\bf 11} (1998) 018,
\href{http://www.arXiv.org/abs/hep-th/9807205}{{hep-th/9807205}}.

\bibitem{Gross:1980he}
D.~J. Gross and E.~Witten, ``POSSIBLE THIRD ORDER PHASE TRANSITION IN THE LARGE
  N LATTICE GAUGE THEORY,'' { Phys. Rev.} {\bf D21} (1980)
446--453.

\bibitem{Wadia:1979vk}
S.~Wadia, ``A STUDY OF U(N) LATTICE GAUGE THEORY IN TWO-DIMENSIONS,''.
  EFI-79/44-CHICAGO.

\bibitem{Wadia:1980cp}
S.~R. Wadia, ``N = infinity PHASE TRANSITION IN A CLASS OF EXACTLY SOLUBLE
  MODEL LATTICE GAUGE THEORIES,'' { Phys. Lett.} {\bf B93} (1980)
403.

\bibitem{Furuuchi:2003sy}
K.~Furuuchi, E.~Schreiber, and G.~W. Semenoff, ``Five-brane thermodynamics from
  the matrix model,''
\href{http://www.arXiv.org/abs/hep-th/0310286}{{hep-th/0310286}}.

\bibitem{Semenoff:2004bs}
G.~W. Semenoff, ``Matrix model thermodynamics,''
\href{http://www.arXiv.org/abs/hep-th/0405107}{{hep-th/0405107}}.

\bibitem{Alvarez-Gaume:2005fv}
L.~Alvarez-Gaume, C.~Gomez, H.~Liu, and S.~Wadia, ``Finite temperature
  effective action, AdS(5) black holes, and 1/N expansion,'' { Phys. Rev.} {\bf
  D71} (2005) 124023,
\href{http://www.arXiv.org/abs/hep-th/0502227}{{hep-th/0502227}}.

\bibitem{'tHooft:1973jz}
G.~'t~Hooft, ``A PLANAR DIAGRAM THEORY FOR STRONG INTERACTIONS,'' { Nucl.
  Phys.} {\bf B72} (1974)
461.

\bibitem{Brigante:2005bq}
M.~Brigante, G.~Festuccia, and H.~Liu, ``Inheritance principle and
  non-renormalization theorems at finite temperature,'' { Phys. Lett.} {\bf
  B638} (2006) 538--545,
\href{http://www.arXiv.org/abs/hep-th/0509117}{{hep-th/0509117}}.

\bibitem{Eguchi:1982nm}
T.~Eguchi and H.~Kawai, ``REDUCTION OF DYNAMICAL DEGREES OF FREEDOM IN THE
  LARGE N GAUGE THEORY,'' { Phys. Rev. Lett.} {\bf 48} (1982)
1063.

\bibitem{Furuuchi:2005eu}
K.~Furuuchi, ``Large N reductions and holography,'' { Phys. Rev.} {\bf D74}
  (2006) 045027,
\href{http://www.arXiv.org/abs/hep-th/0506183}{{hep-th/0506183}}.

\bibitem{Berezinsky:1970pd}
V.~L. Berezinski { Ph.D Thesis (1970)}.

\bibitem{Berezinsky:1972aa}
V.~L. Berezinsky { Sov. Phys. JETP} {\bf 34} (1972)
610.

\bibitem{Kosterlitz:1973xp}
J.~M. Kosterlitz and D.~J. Thouless, ``Ordering, metastability and phase
  transitions in two- dimensional systems,'' { J. Phys.} {\bf C6} (1973)
1181--1203.

\bibitem{Polyakov:1987ez}
A.~M. Polyakov, ``GAUGE FIELDS AND STRINGS,'' { CHUR, SWITZERLAND: HARWOOD
  (1987) 301 P. (CONTEMPORARY CONCEPTS IN PHYSICS, 3)}.

\bibitem{Atick:1988si}
J.~J. Atick and E.~Witten, ``THE HAGEDORN TRANSITION AND THE NUMBER OF DEGREES
  OF FREEDOM OF STRING THEORY,'' { Nucl. Phys.} {\bf B310} (1988)
291--334.

\bibitem{Sathiapalan:1986db}
B.~Sathiapalan, ``VORTICES ON THE STRING WORLD SHEET AND CONSTRAINTS ON TORAL
  COMPACTIFICATION,'' { Phys. Rev.} {\bf D35} (1987)
3277.

\bibitem{Kogan:1987jd}
Y.~I. Kogan, ``VORTICES ON THE WORLD SHEET AND STRING'S CRITICAL DYNAMICS,'' {
  JETP Lett.} {\bf 45} (1987)
709--712.

\bibitem{Kazakov:2000pm}
V.~Kazakov, I.~K. Kostov, and D.~Kutasov, ``A matrix model for the
  two-dimensional black hole,'' { Nucl. Phys.} {\bf B622} (2002) 141--188,
\href{http://www.arXiv.org/abs/hep-th/0101011}{{hep-th/0101011}}.

\bibitem{Alexandrov:2001cm}
S.~Alexandrov and V.~Kazakov, ``Correlators in 2D string theory with vortex
  condensation,'' { Nucl. Phys.} {\bf B610} (2001) 77,
\href{http://www.arXiv.org/abs/hep-th/0104094}{{hep-th/0104094}}.

\bibitem{Alexandrov:2003ut}
S.~Alexandrov, ``Matrix quantum mechanics and two-dimensional string theory in
  non-trivial backgrounds,''
\href{http://www.arXiv.org/abs/hep-th/0311273}{{hep-th/0311273}}.

\bibitem{Gross:1990md}
D.~J. Gross and I.~R. Klebanov, ``Vortices and the nonsinglet sector of the c =
  1 matrix model,'' { Nucl. Phys.} {\bf B354} (1991)
459--474.

\bibitem{Gross:1990ub}
D.~J. Gross and I.~R. Klebanov, ``ONE-DIMENSIONAL STRING THEORY ON A CIRCLE,''
  { Nucl. Phys.} {\bf B344} (1990)
475--498.

\bibitem{Boulatov:1991xz}
D.~Boulatov and V.~Kazakov, ``One-dimensional string theory with vortices as
  the upside down matrix oscillator,'' { Int. J. Mod. Phys.} {\bf A8} (1993)
  809--852,
\href{http://www.arXiv.org/abs/hep-th/0012228}{{hep-th/0012228}}.

\bibitem{Suyama:2004vk}
T.~Suyama and P.~Yi, ``A holographic view on matrix model of black hole,'' {
  JHEP} {\bf 02} (2004) 017,
\href{http://www.arXiv.org/abs/hep-th/0401078}{{hep-th/0401078}}.

\bibitem{Basu:2005pj}
P.~Basu and S.~R. Wadia, ``R-charged AdS(5) black holes and large N unitary
  matrix models,'' { Phys. Rev.} {\bf D73} (2006) 045022,
\href{http://www.arXiv.org/abs/hep-th/0506203}{{hep-th/0506203}}.

\bibitem{Hori:2001ax}
K.~Hori and A.~Kapustin, ``Duality of the fermionic 2d black hole and N = 2
  Liouville theory as mirror symmetry,'' { JHEP} {\bf 08} (2001) 045,
\href{http://www.arXiv.org/abs/hep-th/0104202}{{hep-th/0104202}}.

\bibitem{McGreevy:2003kb}
J.~McGreevy and H.~L. Verlinde, ``Strings from tachyons: The c = 1 matrix
  reloaded,'' { JHEP} {\bf 12} (2003) 054,
\href{http://www.arXiv.org/abs/hep-th/0304224}{{hep-th/0304224}}.

\bibitem{Takahasi:1974zn}
Y.~Takahasi and H.~Umezawa, ``Thermo field dynamics,'' { Collect. Phenom.} {\bf
  2} (1975)
55--80.

\bibitem{Umezawa:1982nv}
H.~Umezawa, H.~Matsumoto, and M.~Tachiki, ``THERMO FIELD DYNAMICS AND CONDENSED
  STATES,''. Amsterdam, Netherlands: North-holland (1982) 591p.

\bibitem{Semenoff:1982ev}
G.~W. Semenoff and H.~Umezawa, ``FUNCTIONAL METHODS IN THERMO FIELD DYNAMICS: A
  REAL TIME PERTURBATION THEORY FOR QUANTUM STATISTICAL MECHANICS,'' { Nucl.
  Phys.} {\bf B220} (1983)
196--212.

\bibitem{Niemi:1983nf}
A.~J. Niemi and G.~W. Semenoff, ``FINITE TEMPERATURE QUANTUM FIELD THEORY IN
  MINKOWSKI SPACE,'' { Ann. Phys.} {\bf 152} (1984)
105.

\bibitem{Landsman:1986uw}
N.~P. Landsman and C.~G. van Weert, ``REAL AND IMAGINARY TIME FIELD THEORY AT
  FINITE TEMPERATURE AND DENSITY,'' { Phys. Rept.} {\bf 145} (1987)
141.

\bibitem{LeBallac:1996bm}
M.~LE~BELLAC, ``Thermal Field Theory,'' { Cambridge Univ. Pr. (1996) 256 p}.

\bibitem{Umezawa:1993yq}
H.~Umezawa, ``Advanced field theory: Micro, macro, and thermal physics,'' { New
  York, USA: AIP (1993) 238 p}.

\bibitem{Ojima:1981ma}
I.~Ojima, ``GAUGE FIELDS AT FINITE TEMPERATURES: 'THERMO FIELD DYNAMICS', KMS
  CONDITION AND THEIR EXTENSION TO GAUGE THEORIES,'' { Ann. Phys.} {\bf 137}
  (1981)
1.

\bibitem{Fidkowski:2003nf}
L.~Fidkowski, V.~Hubeny, M.~Kleban, and S.~Shenker, ``The black hole
  singularity in AdS/CFT,'' { JHEP} {\bf 02} (2004) 014,
\href{http://www.arXiv.org/abs/hep-th/0306170}{{hep-th/0306170}}.

\bibitem{Balasubramanian:1998de}
V.~Balasubramanian, P.~Kraus, A.~E. Lawrence, and S.~P. Trivedi, ``Holographic
  probes of anti-de Sitter space-times,'' { Phys. Rev.} {\bf D59} (1999)
  104021,
\href{http://www.arXiv.org/abs/hep-th/9808017}{{hep-th/9808017}}.

\bibitem{Horowitz:1998xk}
G.~T. Horowitz and D.~Marolf, ``A new approach to string cosmology,'' { JHEP}
  {\bf 07} (1998) 014,
\href{http://www.arXiv.org/abs/hep-th/9805207}{{hep-th/9805207}}.

\bibitem{Maldacena:2001kr}
J.~M. Maldacena, ``Eternal black holes in Anti-de-Sitter,'' { JHEP} {\bf 04}
  (2003) 021,
\href{http://www.arXiv.org/abs/hep-th/0106112}{{hep-th/0106112}}.

\bibitem{Herzog:2002pc}
C.~P. Herzog and D.~T. Son, ``Schwinger-Keldysh propagators from AdS/CFT
  correspondence,'' { JHEP} {\bf 03} (2003) 046,
\href{http://www.arXiv.org/abs/hep-th/0212072}{{hep-th/0212072}}.

\bibitem{Baier:1993yh}
R.~Baier and A.~Niegawa, ``Analytic continuation of thermal N point functions
  from imaginary to real energies,'' { Phys. Rev.} {\bf D49} (1994) 4107--4112,
\href{http://www.arXiv.org/abs/hep-ph/9307362}{{hep-ph/9307362}}.

\bibitem{Banks:1998dd}
T.~Banks, M.~R. Douglas, G.~T. Horowitz, and E.~J. Martinec, ``AdS dynamics
  from conformal field theory,''
\href{http://www.arXiv.org/abs/hep-th/9808016}{{hep-th/9808016}}.

\bibitem{Schwinger:1960qe}
J.~S. Schwinger, ``Brownian motion of a quantum oscillator,'' { J. Math. Phys.}
  {\bf 2} (1961)
407--432.

\bibitem{Keldysh:1964ud}
L.~V. Keldysh, ``Diagram technique for nonequilibrium processes,'' { Zh. Eksp.
  Teor. Fiz.} {\bf 47} (1964)
1515--1527.

\bibitem{Mathur:1993tp}
S.~D. Mathur, ``Is the Polyakov path integral prescription too restrictive?,''
\href{http://www.arXiv.org/abs/hep-th/9306090}{{hep-th/9306090}}.

\bibitem{Kraus:2002iv}
P.~Kraus, H.~Ooguri, and S.~Shenker, ``Inside the horizon with AdS/CFT,'' {
  Phys. Rev.} {\bf D67} (2003) 124022,
\href{http://www.arXiv.org/abs/hep-th/0212277}{{hep-th/0212277}}.

\bibitem{Festuccia:2005pi}
G.~Festuccia and H.~Liu, ``Excursions beyond the horizon: Black hole
  singularities in Yang-Mills theories. I,'' { JHEP} {\bf 04} (2006) 044,
\href{http://www.arXiv.org/abs/hep-th/0506202}{{hep-th/0506202}}.

\bibitem{Susskind:1993if}
L.~Susskind, L.~Thorlacius, and J.~Uglum, ``The Stretched horizon and black
  hole complementarity,'' { Phys. Rev.} {\bf D48} (1993) 3743--3761,
\href{http://www.arXiv.org/abs/hep-th/9306069}{{hep-th/9306069}}.

\bibitem{Jacobson:1994fp}
T.~Jacobson, ``A Note on Hartle-Hawking vacua,'' { Phys. Rev.} {\bf D50} (1994)
  6031--6032,
\href{http://www.arXiv.org/abs/gr-qc/9407022}{{gr-qc/9407022}}.

\bibitem{Wald:1995yp}
R.~M. Wald, ``Quantum field theory in curved space-time and black hole
  thermodynamics,'' { Chicago Univ. Pr. (1994) 205 p}.

\bibitem{Israel:1976ur}
W.~Israel, ``Thermo field dynamics of black holes,'' { Phys. Lett.} {\bf A57}
  (1976)
107--110.

\bibitem{Maldacena:2005hi}
J.~M. Maldacena, ``Long strings in two dimensional string theory and non-
  singlets in the matrix model,'' { JHEP} {\bf 09} (2005) 078,
\href{http://www.arXiv.org/abs/hep-th/0503112}{{hep-th/0503112}}.

\end{thebibliography}\endgroup
\bibliographystyle{kazu}

\end{document}